\documentclass[11pt]{elsarticle}

\usepackage{textcomp}

\usepackage{hyperref}

\usepackage[a4paper, left=2cm,right=2cm,top=2cm,bottom=2cm]{geometry}

\usepackage{float}
\usepackage{subcaption}
\usepackage{relsize}

\usepackage{siunitx}
\usepackage{bm}
\usepackage{amsmath}
\usepackage{mathrsfs}
\usepackage{lscape}

\usepackage{booktabs} 

\usepackage{graphicx}
\usepackage{amssymb}
\usepackage{amsthm}
\usepackage{graphicx}

\usepackage{tikz}
\usetikzlibrary{plotmarks}
\usetikzlibrary{spy}
\usepackage{pgfplots}
\usetikzlibrary{matrix,shapes,arrows,positioning,chains}
\tikzset{
    decision/.style={
        diamond,
        draw,
        aspect=2,
        text width=10em,
        text badly centered,
        inner sep=0pt
    },
    block/.style={
        rectangle,
        draw,
        text width=10em,
        text centered,
        rounded corners
    },
    cloud/.style={
        draw,
        ellipse,
        minimum height=2em
    },
    descr/.style={
        fill=white,
        inner sep=2.5pt
    },
    connector/.style={
        -latex,
        font=\scriptsize
    },
    rectangle connector/.style={
        connector,
        to path={(\tikztostart) -- ++(#1,0pt) \tikztonodes |- (\tikztotarget) },
        pos=0.5
    },
    rectangle connector/.default=-2cm,
    straight connector/.style={
        connector,
        to path=--(\tikztotarget) \tikztonodes
    }
}

\usepackage{natbib}
\usepackage{lineno}
\modulolinenumbers[5]
\usepackage{tgpagella}
\usepackage{newpxmath}
\usepackage{cleveref}

\hypersetup{
	colorlinks,
	pdfborder={0 0 0},
	linkcolor={blue!50!black},
	citecolor={blue!50!black},
	urlcolor={blue!80!black},
	anchorcolor = {blue!80!black},
	filecolor = {blue!80!black},
	menucolor = {blue!80!black},
	runcolor = {blue!80!black}
}

\newcommand{\norm}[1]{\left\lVert#1\right\rVert}

\renewcommand{\vec}[1]{\mathbf{#1}}

\newtheorem{definition}{Definition}
\newtheorem{remark}{Remark}

\journal{}

\begin{document}
\begin{frontmatter}



\title{Computational framework for resolving boundary layers in electrochemical systems using weak imposition of Dirichlet boundary conditions}


\author[1,2]{Sungu Kim\fnref{SunguFootnote}}
\author[1,3]{Makrand A. Khanwale\fnref{MakFootnote}}
\author[2]{Robbyn K. Anand}
\author[1]{Baskar Ganapathysubramanian}
\address[1]{Department of Mechanical Engineering, Iowa State University, 2025 Black Engineering, Ames, IA 50011, USA}
\address[2]{Department of Chemistry, Iowa State University, 1605 Gilman Hall, Ames, IA 50011, USA}
\address[3]{Department of Mathematics, Iowa State University, Iowa, USA 50011}
\fntext[SunguFootnote]{Presently at Department of Mechanical Engineering, Stanford University, CA, USA}
\fntext[MakFootnote]{Presently at Center for Turbulence Research, Department of Mechanical Engineering, Stanford University, CA, USA}

\begin{abstract}
We present a finite element based computational framework to model electrochemical systems.  The electrochemical system is represented by the coupled Poisson-Nernst-Planck (PNP) and Navier-Stokes (NS) equations. The key quantity of interest in such simulations is the current (flux) at the system boundaries. Accurately computing the current flux is challenging due to the small critical dimension of the boundary layers (small Debye layer) that require fine mesh resolution at the boundaries.  
We present a numerical framework which resolves this challenge by utilizing a weak imposition of Dirichlet boundary conditions for Poisson-Nernst-Plank equations.  In this numerical framework we utilize a block iterative strategy to solve NS and PNP equations.  This allows us to efficiently and easily implement the weak imposition of Dirichlet boundary conditions.  
The results from our numerical framework shows excellent agreement when compared to strong imposition of boundary conditions (strong imposition requires a much finer mesh).  
Furthermore, we show that the weak imposition of the boundary conditions allows us to resolve the fluxes in the boundary layers with much coarser meshes compared to strong imposition.  We also show that the method converges as we refine the mesh near the boundaries at a much faster rate compared to strong imposition of the boundary layer.   We present multiple test cases with varying boundary layer thickness to illustrate the utility of the numerical framework.  We illustrate the approach on canonical 3D problems that otherwise would have been computationally intractable to solve accurately. Lastly, we simulate electrokinetic instabilities near a perm-selective membrane with weakly imposed boundary conditions on the membrane. This approach substantially reduces the computational cost of modeling thin boundary layers in electrochemical systems. 
\end{abstract}


\end{frontmatter}

\section{Introduction}
\label{intro}
An understanding of charged species transport is critical to the development of electrochemical and electrokinetic systems relevant to a wide range of disciplines (engineering, chemistry, physics) and applications (sensing, energy, water purification). Systems that employ non-linear electrokinetics, in which the electric field is varied spatially (and also temporally in some cases), are especially difficult to model. For example, charged species can be electrokinetically focused along a steep electric field gradient formed near an ion-selective membrane or a bipolar electrode (BPE)~\cite{li2016recent,mavre2010bipolar}. In both cases, the electric field gradient results from the local depletion of charge carriers at one end of the membrane (by selective charge transport) or BPE (by faradaic reactions)~\cite{bondarenko2020current}. The formation of an ion depletion zone (IDZ) and ion enrichment zone (IEZ) at opposite sides of the membrane or BPE is called ion concentration polarization (ICP). A few prominent applications include water purification and desalting~\cite{kim2010direct,berzina2018electrokinetic}, biomedical engineering~\cite{berzina2018electrokinetic}, and enrichment and detection of trace analytes~\cite{anand2011bipolar,kim2020concentration}. In all of these applications, the stability of the IDZ drastically limits the volumetric throughput of these devices. Therefore, the ability to simulate species transport in these systems is critical to their advancement.

Species transport in electrochemical systems, such as ICP, is a complex multi-physics problem driven by diffusion, electromigration, and convection \cite{probstein2005physicochemical}. Experimental approaches employed to characterize this multi-physics problem are generally limited to the measurement of electrical current or to the visualization of fluorescent tracer molecules. As a result, it is difficult to fully understand the mechanism of ICP using these methods alone. Therefore, there have been several numerical studies of transport in such systems to compliment experimental results. For example, Zangle et al.~\cite{mani2009propagation,zangle2009propagation} derived 1-dimensional (1D) governing equations for ICP in a system comprising a micro-nano-micro junction and calculated shock wave-like IDZ and IEZ propagation along the microchannel segments, originating at the nanochannel. Using this approach, they found that Dukhin number (surface conductivity over bulk fluid conductivity) and the electrophoretic mobility of charged species are the parameters that dictate the rate and extent of the propagation. Numerical results obtained by Mani and collaborators showed that chaotic fluid motion originates from the locally high electric field~\cite{druzgalski2013direct} or alternating current (AC)~\cite{kim2019characterization} even in the low Reynolds number regime. ICP is made further complex when, in addition to convection, diffusion, and migration, chemical reactions are involved. To address such a case, Kler et al. included reaction terms to simulate electrophoresis accompanied by acid and base reactions \cite{kler2011modeling}. Similarly, Tallarek and coworkers included acid/base and faradaic reaction terms in the simulation of ICP at BPEs \cite{hlushkou2016numerical}.

Although these studies exemplify successful simulation of charged species transport in non-linear electrokinetics, it is still challenging to obtain reliable results with a reasonable computational cost. The primary reason for this difficulty is the multi-scale nature of the problem~\cite{boy2008simulation}. The smallest scale feature that impacts the physics in electrochemical and electrokinetic systems is the electrical double layer ($\sim$\(10 nm\)) or EDL, which comprises electrical potential and ion concentration gradients in the boundary layer present at a liquid-solid interface. In contrast, species transport relevant to most applications of such systems extend over length scales of $\sim$\(10 \mu m\) to $\sim$\(1000 \mu m\). Importantly, unresolved boundary layers can result in unfavorable oscillations extending outside of the boundary layers into the bulk domain. Refining the mesh near the boundary is a reasonable approach to address challenges from multi-scale characteristics \cite{druzgalski2013direct,patankar1998numerical} and provides reliable results in the entire domain. However, the computational cost incurred by the increased mesh density in the boundary layer can be prohibitive for very small Debye lengths. Considering that most applications are interested in what happens in the 'bulk' of the fluid domain (and its impact on current flux), not in the vicinity of the boundary, resolving the mesh near this layer is not computationally economical. Jia et al. ~\cite{jia2014multiphysics1,jia2014multiphysics2}, using a commercial code, simplified complex boundary physics with electroosmotic slip velocity, which minimizes computational costs. However, there is ambiguity in the selection of the location where the slip boundary condition imposed away from an ion selective membrane transitions to a no-slip boundary condition, imposed on or adjacent to the membrane. Moreover, replacing the boundary layer with the slip boundary condition ignores concentration gradient driven flow \cite{alizadeh2017multiscale,cho2014overlimiting}. Therefore, there remains a need to reduce the computational cost of representing the boundary layer without oversimplifying the underlying physics.  

In this work, we address this need by utilizing an approach used in fluid mechanics -- the Dirichlet-to-Neumann transformation -- this is used to efficiently model the no-slip condition (Dirichlet boundary condition)\cite{bazilevs2007weak}. The Dirichlet-to-Neumann transformation, also known as Nitsche's method or symmetric interior penalty Galerkin method (SIPG), provides a consistent and robust way of enforcing Dirichlet conditions by variational weakening of the no-slip condition into a Neumann type condition, especially in the context of Finite Element (FE) analysis. \cite{juntunen2009nitsche, brenner2008weakly} Such a strategy releases the point-wise no-slip condition imposed at the boundary of the fluid domain, thus minimizing the mesh resolution required to track the steep gradients close to the boundaries. This effect reliably imitates the presence (and effect) of the thin boundary layer~\cite{bazilevs2009computational,hsu2012wind}. Enforcing Dirichlet boundary conditions weakly allows for an accurate overall flow solution even if the mesh size in the wall-normal direction is relatively large. This approach has substantially benefited efficient simulations of turbulent flow scenarios~\cite{bazilevs2007turbulent,bazilevs2007variational} as well as other multi-physics flow scenarios~\cite{hansbo2003nitsche,xu2019residual}. 

%
The present study develops a FEM framework for the fully coupled Navier-Stokes and Poisson-Nernst-Planck (NS-PNP) equations to simulate electrochemical and electrokinetic systems. To overcome difficulties from thin boundary layers, Dirichlet boundary conditions are weakly enforced \cite{nitsche1971variationsprinzip, bazilevs2007weak} in the PNP equations. While usage of the developed framework is not limited to specific application to electrochemical and electrokinetic systems, we demonstrate its efficacy in resolving calculations of electroosmotic flow (EOF) and ion concentration polarization (ICP). EOF and ICP were selected as test cases for two reasons - first, there has been growing interest in these phenomena due to their potential impact in chemical, biomedical, and environmental fields, and second, because these examples include the three fundamental transport mechanisms - convection, diffusion, and migration. Our findings are significant because, despite a much coarser mesh, the results obtained with weak BC showed good agreement to those obtained with strong BC, and furthermore, boundary flux calculations converged much faster to the solution using the weak BC. Collectively, these results demonstrate a significant reduction in computational load while retaining accuracy. Therefore, we expect that this weak BC approach will provide greater stability and accuracy in the simulation of a wide range of electrochemical and electrokinetic systems.

The outline of the rest of the paper is as follows: We begin by revisiting the governing equations for charged species transport followed by the non-dimensional forms of these governing equations in Section 2. Then, the FEM problems are defined with weakly imposed Dirichlet boundary conditions in Section \ref{Problem formulation}. The solving strategy for the numerical method was discuss in Section \ref{section:NumericalMethod}. In Section \ref{sec:results}, the developed framework is validated with a manufactured solution and by simulation of EOF, for which an analytical solution exists. Next, 1D and 2D IDZs are simulated with weakly imposed Dirichlet boundary conditions. For 3D applications, the developed platform was tested for electrolyte separation (desalting) in a microchannel. 
For each example, the calculated boundary flux is compared with that obtained with strongly imposed Dirichlet boundary conditions. 
We also test the weakly imposed boundary conditions for the simulation of the electrokinetic instabilities near a perm-selective membrane.
We conclude in Section \ref{sec:conclusion}.
\section{Charged species transport}\label{chargedspeciestransport}
\subsection{Governing equations} \label{GoverningEquations}
\noindent\textbf{Poisson-Nernst-Planck (PNP)}: Without loss of generality, we consider a canonical problem of solvent flow and species transport in a (micro)channel configuration. This problem encompasses both the electroosmotic and pressure driven regimes. We consider $N > 1$ number of charged species with subscript \(i\) indicating the species index. The species flux $\bm{j}_i^*$\footnote{The asterisk (*) is used for dimensional quantities, so that the notation is simplified when we consider non-dimensional terms.}, which is driven by diffusion, migration, and convection, is written as:

\begin{equation}
	\label{eq:NPFlx}
    \vec{j_{i}}^* = -D_i\nabla^* c_{i}^* - D_i\frac{z_{i}F}{RT}c_{i}^*\nabla^* \phi^* + \vec{u}^* c_{i}^*.
\end{equation}
Eq.~(\ref{eq:NPFlx}) is the Nernst-Planck equation~\cite{probstein2005physicochemical} for the $i^{th}$ species. \(D_i\) is the diffusivity, \(c_i^*\) is concentration of the species, \(z_i\) is the valence of species, \(F\) is the Faraday constant, 
\(R\) is the gas constant, 
\(T\) is the temperature, \(\phi^*\) is the electric potential, and \(\vec{u}^*\)\footnote{We follow the convention that bold symbols represent vectors with dimension, $d$. } is the fluid velocity. We get the rate of change of the species concentration by considering flux balance,

\begin{equation}
\label{eq:NPDimensional}
\frac{\partial c_i^*}{\partial t^*}+ \vec{u}^* \cdot \nabla^* c_{i}^*
    = \nabla^* \cdot (D_i\nabla^* c_{i}^* + D_i\frac{z_{i}F}{RT}c_{i}^*\nabla^* \phi^*). 
\end{equation}
Potential \(\phi^*\) is obtained from the Poisson equation, which describes Gauss's law,

\begin{equation}\label{eq:PoissonDimensional}
    -\varepsilon\nabla^{*2}\phi^* = \rho_e^*, 
\end{equation}
where \(\varepsilon\) is the electric permittivity of solvent, and \(\rho_e^*\) is the charge density given by 
\begin{equation}\label{eq:chargeDensity}
    \rho_e^*=F\sum_{i=1}^N z_i^*c_i^*.
\end{equation}

\noindent \textbf{Boundary conditions (PNP)}: We focus on the boundary conditions of a permselective membrane. Typical boundary conditions for the counter-ion concentration $c_i^*$ at the permselective membrane are Dirichlet \begin{equation}\label{eq:charge_DBC}
    c_i^* = c_{i,M}^*
\end{equation}
and (zero) Neumann for co-ion species
\begin{equation}\label{eq:charge_NBC}
    \vec{j_{i}}^*\cdot\vec{n} = 0
\end{equation}
where, $\vec{n}$ is the outward pointing normal.  
The boundary conditions for the potential at the permselective membrane are also typically Dirichlet
\begin{equation}\label{eq:phi_DBC}
    \phi^* = \phi_M^*
\end{equation}

\begin{remark}
	The zero Neumann condition Eq.~\ref{eq:charge_NBC} represents zero flux across the boundary. To maintain a zero current flux, diffusion and electric migration (in Eq.~\ref{eq:NPFlx}) cancel each other at the boundary. As a result, there can be non-zero gradients of the concentration and the potential with a zero current flux boundary condition. This condition is in contrast to heat transfer or diffusion-convection problems involving a single variable.
\end{remark}

\noindent \textbf{Navier-Stokes (NS)}: In conjunction with the Poisson-Nernst-Planck equations, the solvent momentum transport is described by the Navier-Stokes equation
\begin{equation}\label{eq:NSDim}
    \rho^*\frac{\partial \vec{u}^*}{\partial t^*}+\nabla^* \cdot (\rho^* \vec{u}^*\otimes \vec{u}^*)=-\nabla^* p^* + \eta^*\nabla^{*2}\vec{u}^*+\vec{f_b}^*.
\end{equation}
\(\rho^*\) is the density of solution, \(p\) is pressure, \(\eta^*\) is the dynamic viscosity. The last term of equation (\ref{eq:NSDim}) is the body force due to an electric field acting on charged species, which couples the Navier-Stokes equation with equations (\ref{eq:NPDimensional}) and (\ref{eq:PoissonDimensional})

\begin{equation}\label{eq:bodyForce}
    \vec{f_b}^*=-F\sum_{i=1}^N c_i^* z_i^*\nabla^* \phi^*.
\end{equation}
The carrier fluid is assumed to be incompressible
\begin{equation}\label{eq:NDContinuity}
    \nabla^* \cdot \vec{u}^* = 0.
\end{equation}

\noindent \textbf{Boundary conditions}: At the electrodes, the no-slip condition for velocity is enforced, $\vec{u^*} = 0$

\subsection{Non-dimensional forms of governing equations} \label{NDequations}

The variables and operators in the governing equations are scaled by characteristic quantities to obtain non-dimensional forms of the governing equations,
\begin{equation} \label{eq:scalingV}
x=\frac{x^*}{L_c}, \quad
\vec{u}=\frac{\vec{u}^*}{U_{c}}, 
\quad p=\frac{p^*}{p_{c}}, \quad c_i=\frac{c_i^*}{c_{c}}, \quad \phi=\frac{\phi^*}{\phi_{c}}, \quad \phi=\frac{\rho^*}{\rho_{c}}, \quad \phi=\frac{\eta^*}{\eta_{c}},
\end{equation}

\noindent where subscript \(c\) denotes characteristic quantities. The reference length \(L_{c}\)\footnote{From now on we drop the subscript \(c\) for concise notation.} is chosen to be the channel width, \(L\). The characteristic concentration, potential, fluid velocity, pressure, and time scale (which is derived from velocity and length references) respectively are as follows, 

\begin{equation} \label{eq:refQ}
\begin{split}
    c_{c}=I_b=\frac{1}{2}\sum_{i=1}^N z_i^2 c_{i}^{initial},
    \quad \phi_{c} = V_T=\frac{RT}{F} \\
    U_{c}=\frac{D}{L}, \quad p_{c}=\frac{\eta D}{L^2}, \quad \tau=\frac{L^2}{D}.
\end{split}
\end{equation}
Where \(I_b\) is the ionic strength of the bulk electrolyte and \(V_T\) is thermal voltage, $D$ is the average diffusion coefficient of the species. Substituting dimensional quantities and operators with normalized variables and operators provides the non-dimensional equations as follows, 




\begin{align}
	\textit{Species flux:} &  \quad \vec{j}_{i} = -\nabla c_{i} - z_{i}c_{i}\nabla \phi + \vec{u} c_{i}, \label{eq:NonDimFlx}\\
	\textit{Nernst-Planck:} &  \quad \frac{\partial c_i}{\partial t} + \vec{u} \cdot \nabla c_{i}
	= \nabla \cdot (\nabla c_{i} + z_{i} c_{i}\nabla \phi), \label{eq:NDNP}\\
	\textit{Poisson:} &  \quad  -2 \Lambda^2 \nabla^2\phi = \rho_e, \label{eq:NDP}\\
	\textit{Navier-Stokes:} 
	&\quad\frac{1}{Sc}\left(\frac{\partial \vec{u}}{\partial t} + \vec{u}\cdot\nabla\vec{u}\right) =-\nabla p + \nabla^{2}\vec{u}
	+\vec{f_b}\label{eq:NDNS}\\
	\textit{Continuity: }&\quad \nabla \cdot \vec{u} = 0, \label{eq:NDCont}\\
	\textit{Normalized charge density:}  &  \quad  \rho_e=\sum_{i=1}^N z_i c_i,\label{eq:chargeDensityND} \\ 
	\textit{Body force:} &  \quad  \vec{f_b}=- \frac{\kappa}{2\Lambda^2}\sum_{i=1}^N c_i z_i\nabla \phi, \label{eq:chargeDensityND}
\end{align}

%
%
%

\noindent where \(\Lambda\) is the normalized Debye length \footnote{The Debye length, or Debye screening length, \(\lambda\) characterizes the electrokinetics near a charged wall. The surface charge at the wall repels co-ions and attracts counter-ions. This electrokinetic repulsion and attraction is countered by thermal energy, thereby forming a diffuse layer adjacent of the wall. \(\lambda\) is the length from the wall into the fluid at which the electric static potential balances the thermal energy \cite{probstein2005physicochemical}. } \(\Lambda = \lambda/L\),

\begin{equation}\label{eq:Debye}
    \lambda = \sqrt{\frac{1}{2}\frac{\varepsilon RT}{F^2I_b}}.
\end{equation}

\noindent \(Sc = \frac{\eta}{\rho D}\) is the Schmidt number which is the ratio of viscous effects to diffusion, and \(\kappa\) is electrohydrodynamic coupling constant~\cite{druzgalski2013direct} given by,
\begin{equation}\label{eq:electrohydrodynamicconst3}
    \kappa = \frac{\varepsilon}{\eta D} \left (\frac{RT}{F}\right)^2.
\end{equation}
and the non-dimensionalized boundary conditions corresponding to those detailed in the previous sub-section.
\\

\begin{remark}
For a typical microchannel (channel hydraulic diameter ranging from 1\(\mu m\) to 1\(mm\)), \(\Lambda\) is small (ranging from \(\Lambda = \) $\num{1e-2}$ to $\num{1e-5}$), forming a thin boundary layer for species concentration and potential. However, the flow is in the laminar regime. Thus, we focus on applying the Dirichlet-to-Neumann transformation only on the PNP equations, and strongly enforce the no-slip conditions for velocity. 
\end{remark}
\begin{remark}
There are several alternate choices for the characteristic length scale, which in turn affect the characteristic timescale. One alternative is to use the Debye length, $\lambda$ as the characteristic length. This results in a very small characteristic timescale~\cite{morrow2006time}. Another alternative defines the characteristic length scale as the harmonic mean of the channel hydraulic diameter and the Debye length, $L_c =  \sqrt{L\lambda}$. These alternative timescales are particularly useful to resolve scenarios with small \(\Lambda\). See details in \ref{app:lambdaT}. In this study, $L_c = \sqrt{L\lambda}$ (and thus, \(\tau = L\lambda/D\)) was used for small \(\Lambda\) (\(\Lambda<\SI{1e-2}{}\)), while $L_c = L$ (and thus, \(\tau = L^2/D\)) was used for moderate to large \(\Lambda\) (\(\Lambda\geq\SI{1e-2}{}\)). 
\end{remark}

\section{Variational form and the Dirichlet-to-Neumann transformation} \label{Problem formulation}
\subsection{Weak form of the equations} 
\label{section:Variational problem and finite element approximation} 
Consider the spatial domain as $\Omega_D$, with $\partial\Omega_D$ as its boundary, and by $\Gamma_D$ the boundary where the weak boundary conditions are enforced.  We can define the variational problem as follows. 



\begin{definition}
	Let $(\cdot,\cdot)$ be the standard $L^2$ inner product over the subscript (i.e. either $\Omega_D$ or $\partial\Omega_D/\Gamma_D$). We state the variational problem as follows: find $\vec{u}(\vec{x}) \in \vec{H}_0^1(\Omega)$, $c_1(\vec{x}),...,c_N(\vec{x}), \phi(\vec{x}), p(\vec{x})$ $\in {H}^1(\Omega)$ such that\footnote{Here the subscript 0 for the Sobolev space $\vec{H}_{0}^1(\Omega)$ represents zero velocities on the boundary in the trace sense.}
	\begin{align}
		\text{Nernst Planck Eqns:} & \quad \mathcal{B}_{NP,i}\Big(q;c_i,\phi,\vec{u}\Big) 
		+ \mathcal{L}_{NP,i}\Big(q;c_i,\phi\Big) = 0, ~\texttt{for}~i = 1,...,N, 
		\label{varNP}\\ 
		\text{Poisson Eqn:} & \quad \mathcal{B}_{P}\Big(q;\phi\Big) 
		+ \mathcal{L}_{P}\Big(q;\phi\Big) = 0, \label{varPoisson} \\
		\text{Navier-Stokes:} & \quad 
		\mathcal{B}_{NS}\Big(\vec{w},q;\vec{u},p\Big) + \mathcal{L}_{NS}\Big(\vec{w};\vec{u}\Big) = 0, \label{varNS}
	\end{align}
	$\forall \vec{w} \in \vec{H}^1_0(\Omega)$, $\forall q \in H^1(\Omega)$.
	\label{def:variational_form}
\end{definition}
\noindent where $\mathcal{B}$ and $\mathcal{L}$ represent the bilinear and linear forms respectively for each equation given by,
\begin{align} 
\begin{split}
    \textit{Nernst-Plank Eqns:}
    &\quad
    \mathcal{B}_{NP,i}\Big(q;c_i,\phi,\vec{u}\Big) = \Big(q,\frac{\partial c_i}{\partial t}\Big)_{\Omega_D} 
    +\Big(q,\vec{u}\cdot \nabla c_i\Big)_{\Omega_D}
    +(\nabla q,\nabla c_i)_{\Omega_D}\\
    &\quad\quad\quad\quad\quad\quad\quad\quad\quad\quad
    +(\nabla q,z_i c_i \nabla \phi)_{\Omega_D},
\end{split}\label{eq:varNPBilinear}\\
\begin{split}
    &\quad \mathcal{L}_{NP,i}\Big(q;c_i,\phi\Big) =
    -(q,\nabla c_i \cdot \vec{n}+z_{i}c_{i} \nabla \phi \cdot \vec{n})_{\partial \Omega_D/\Gamma_D},
\end{split}\label{eq:varNPFlux}\\
\begin{split}
	\textit{Poisson Eqns:}
	&\quad
	\mathcal{B}_{P}\Big(q;\phi\Big) = 
	2\Lambda^2(\nabla q; \nabla \phi)_{\Omega_D},
\end{split}\label{eq:varPBilinear}\\
&\quad \mathcal{L}_{P}\Big(r;\phi\Big) = 
-2\Lambda^2(r,\nabla \phi \cdot \vec{n})_{\partial \Omega_D/\Gamma_D}
-(r,\rho_e)_{\Omega_D},\label{eq:varPFlux}\\
\begin{split}
	\textit{Navier-Stokes Eqns:}
	&\quad 
	\mathcal{B}_{NS}\Big(\vec{w},q;\vec{u},p\Big) =
	\frac{1}{Sc}\Big(\vec{w},\frac{\partial\vec{u}}{\partial{t}}\Big)_{\Omega_D}
	+\frac{1}{Sc}\Big(\vec{w},\vec{u}\cdot\nabla\vec{u}\Big)_{\Omega_D}
	+(q,\nabla\cdot\vec{u})_{\Omega_D}\\
	&\quad\quad\quad\quad\quad\quad\quad\quad\quad
	-(\nabla\cdot \vec{w},p)_{\Omega_D}
	+(\nabla{\vec{w}},\nabla{\vec{u}})_{\Omega_D},
\end{split}\label{eq:varNSBilinear}\\
&\quad \mathcal{L}_{NS}\Big(\vec{w};\vec{u},p\Big) = 
-(\vec{w},\nabla\vec{u}\cdot\vec{n})_{\partial \Omega_D}
+(\vec{w}\cdot\vec{n},p)_{\partial \Omega_D}
-(\vec{w},\vec{f_b})_{\Omega_D}.\label{eq:varNSFlx}
\end{align}


\subsection{Semi-discrete time-scheme}\label{section:timescheme}
We utilize a fully-implicit first order backward Euler scheme.  Let $k$ be a time-step; let $t^n := nk$; We can then define the time-discrete variational problem as follows. 
\begin{definition}[time-scheme]
	Let $(\cdot,\cdot)$ be the standard $L^2$ inner product over the subscript (i.e. either $\Omega_D$ or $\partial\Omega_D/\Gamma_D$). We state the variational problem as follows: find $\vec{u}^{n+1}(\vec{x}) \in \vec{H}_0^1(\Omega)$, $c_1^{n+1}(\vec{x}),...,c_N^{n+1}(\vec{x})$, $\phi^{n+1}(\vec{x})$, $p^{n+1}(\vec{x})$ $\in {H}^1(\Omega)$ such that
	\begin{align}
		\text{Nernst Planck Eqns:} & \quad \mathcal{B}_{NP,i}\Big(q;c_i^{n+1},c_i^{n},\phi^{n+1},\vec{u}^{n+1}\Big) 
		+ \mathcal{L}_{NP,i}\Big(q;c_i^{n+1},\phi^{n+1}\Big) = 0, ~\text{for}~i = 1,...,N, 
		\label{varNPtd}\\ 
		\text{Poisson Eqn:} & \quad \mathcal{B}_{P}\Big(q;\phi^{n+1}\Big) 
		+ \mathcal{L}_{P}\Big(q;\phi^{n+1}\Big) = 0, \label{varPoissontd} \\
		\text{Navier-Stokes:} & \quad 
		\mathcal{B}_{NS}\Big(\vec{w},q;\vec{u}^{n+1},\vec{u}^{n}, p^{n+1}\Big) + \mathcal{L}_{NS}\Big(\vec{w};\vec{u}^{n+1},p^{n+1}\Big) = 0, \label{varNStd}
	\end{align}
	$\forall \vec{w} \in \vec{H}^1_0(\Omega)$, $\forall q \in H^1(\Omega)$, given $\vec{u}^{n} \in \vec{H}_0^1(\Omega)$, and $c_1^{n}(\vec{x}),...,c_N^{n}(\vec{x}), \in H^1(\Omega)$.
	\label{def:variational_form_sem_disctd}
\end{definition}
\noindent with the bilinear and linear forms for each equation given by,
\begin{align} 
	\begin{split}
		\textit{Nernst-Plank Eqns:}
		&\quad
		\mathcal{B}_{NP,i}\Big(q;c_i^{n+1},c_i^{n},\phi^{n+1},\vec{u}^{n+1}\Big) = \Big(q,\frac{c_i^{n+1} - c_i^{n}}{k}\Big)_{\Omega_D} 
		+\Big(q,\vec{u}^{n+1}\cdot \nabla c_i^{n+1}\Big)_{\Omega_D}\\
		&\quad\quad\quad\quad\quad\quad\quad\quad\quad\quad\quad\quad\quad\quad
		+\left(\nabla q,\nabla c_i^{n+1}\right)_{\Omega_D}
		+\left(\nabla q,z_i c_i^{n+1} \nabla \phi^{n+1}\right)_{\Omega_D},
	\end{split}\label{eq:varNPBilinear_td}\\
	\begin{split}
		&\quad \mathcal{L}_{NP,i}\Big(q;c_i^{n+1},\phi^{n+1}\Big) =
		-\left(q,\nabla c_i^{n+1} \cdot \vec{n} + z_{i} c_i^{n+1} \nabla \phi^{n+1} \cdot \vec{n}\right)_{\partial \Omega_D/\Gamma_D},
	\end{split}\label{eq:varNPFlux_td}\\
	\begin{split}
		\textit{Poisson Eqns:}
		&\quad
		\mathcal{B}_{P}\Big(q;\phi^{n+1}\Big) = 
		2\Lambda^2\left(\nabla q; \nabla \phi^{n+1}\right)_{\Omega_D},
	\end{split}\label{eq:varPBilinear_td}\\
	&\quad \mathcal{L}_{P}\Big(r;\phi^{n+1}\Big) = 
	-2\Lambda^2\left(r,\nabla \phi^{n+1} \cdot \vec{n}\right)_{\partial \Omega_D/\Gamma_D}
	-\left(r,\rho_e^{n+1}\right)_{\Omega_D},\label{eq:varPFlux_td}\\
	\begin{split}
		\textit{Navier-Stokes Eqns:}
		&\quad 
		\mathcal{B}_{NS}\Big(\vec{w},q;\vec{u}^{n+1},\vec{u}^{n}, p^{n+1}\Big) =
		\frac{1}{Sc}\Big(\vec{w},\frac{\vec{u}^{n+1}-\vec{u}^{n}}{k}\Big)_{\Omega_D}
		+\frac{1}{Sc}\Big(\vec{w},\vec{u}^{n+1}\cdot\nabla\vec{u}^{n+1}\Big)_{\Omega_D}\\
		&\quad\quad\quad\quad\quad\quad\quad\quad\quad\quad\quad\quad\quad
		-\left(\nabla\cdot \vec{w},p^{n+1}\right)_{\Omega_D} 
		+\left(\nabla{\vec{w}},\nabla{\vec{u}^{n+1}}\right)_{\Omega_D}\\
		&\quad\quad\quad\quad\quad\quad\quad\quad\quad\quad\quad\quad\quad
		+\left(q,\nabla\cdot\vec{u}^{n+1}\right)_{\Omega_D}\,\,\, ,
	\end{split}\label{eq:varNSBilinear_td}\\
	&\quad \mathcal{L}_{NS}\Big(\vec{w};\vec{u}^{n+1},p^{n+1}\Big) = 
	-\left(\vec{w},\nabla\vec{u}^{n+1}\cdot\vec{n}\right)_{\partial \Omega_D}
	+\left(\vec{w}\cdot\vec{n},p^{n+1}\right)_{\partial \Omega_D}
	-\left(\vec{w},\vec{f_b}^{n+1}\right)_{\Omega_D}.\label{eq:varNSFlx_td}
\end{align}
Note that the continuity equation is combined with the Navier-Stokes equation here. This is done because this is a fully-implicit pressure coupled time discretisation.  
\subsection{Spatial discretization with stabilization}\label{section:SUPG}
For notational simplicity we consider the time derivatives as continuous while we describe the spatial discretization.  As we seek a continuous Galerkin discretization with equal order interpolation for velocity and pressure, we utilize a popular stabilization-based approach --- streamwise/upwind Petrov--Galerkin (SUPG) in conjunction with pressure stabilized Petrov-Galerkin (PSPG)~\citep{article:BrooksHughes1982, article:TezMitRayShi92}.  The SUPG stabilization also allows us to stabilize advective terms in Navier-Stokes and Nernst-Planck equations.

To achieve the spatial discretization, we substitute the infinite-dimensional spaces in \textbf{\cref{def:variational_form}} by their discrete counterparts (denoted here by a superscript $h$) using conforming Galerkin finite elements augmented along with  SUPG stabilization. Considering a tessellation of the domain $\Omega = \bigcup_{i=1}^{N_{el}} \Omega_i $ into $N_{el}$ non-overlapping elements, the space-discrete form of the Navier-Stokes---Possion-Nernst-Plank (NS-PNP) variational problem is given by:

\begin{definition}
	find $\vec{u}(\vec{x}) \in \vec{H}_0^{1,h}(\Omega)$, $c_1(\vec{x}),...,c_N(\vec{x}), \phi(\vec{x}), p(\vec{x})$ $\in {H}^{1,h}(\Omega)$ such that
	\begin{align}
	\begin{split}
	\text{Nernst Planck Eqns:} & \quad	\mathcal{B}_{NP,i}^{h}\Big(q_i^{h};c_i^{h},\phi^{h},\vec{u}^{h}\Big) 
		+ \mathcal{L}_{NP,i}^{h}\Big(q_i^{h};c_i^{h},\phi^{h}\Big) \\
		\quad\quad\quad\quad\quad\quad\quad\quad\quad\quad
		&+ \textcolor{blue}{\sum_{K=1}^{N_{el}}\Big(\tau_{SUPG}(\nabla\phi^{h} + \vec{u}^{h}) \cdot\nabla q_i^{h},
			\frac{\partial c_i^h}{\partial t} + \vec{u}^h \cdot \nabla c_{i}^h\Big)} = 0, ~\text{for}~i = 1,...,N, 
	\end{split}\label{eq:FEMNP}\\
	\text{Poisson Eqn:}& \quad 
	\mathcal{B}_{P}^{h}\Big(q^{h};\phi^{h}\Big) 
	+ \mathcal{L}_{P}^{h}\Big(q^{h};\phi^{h}\Big) = 0,\label{FEMPoisson}\\
	\begin{split}
		\text{Navier-Stokes Eqns:}& \quad
		\mathcal{B}_{NS}^{h}\Big(\vec{w}^{h},q^{h};\vec{u}^{h},p^{h}\Big) 
		+ \mathcal{L}_{NS}^{h}\Big(\vec{w}^{h},\vec{u}^{h}\Big)\\
		& \quad\quad\quad\quad\quad\quad\quad\quad\quad\quad
		+ \textcolor{blue}{\sum_{K=1}^{N_{el}}\Big(\tau_{SUPG}\vec{u}^{h}\cdot\nabla \vec{w}^{h},
			\frac{\partial \vec{u}^{h}}{\partial t} + \vec{u}^{h}\cdot \nabla \vec{u}^{h} - \vec{f_b}^{h}\Big)}\\
		& \quad\quad\quad\quad\quad\quad\quad\quad\quad\quad
		+ \textcolor{blue}{\sum_{K=1}^{N_{el}}\Big(\tau_{PSPG}\nabla{q}^{h},
			\frac{\partial \vec{u}^{h}}{\partial t} + \vec{u}^{h}\cdot \nabla \vec{u}^{h} - \vec{f_b}^{h} \Big)}= 0,
	\end{split}\label{eq:FEMNS}
\end{align}
$\forall \vec{w} \in \vec{H}^{1,h}_0(\Omega)$, $\forall q \in H^{1,h}(\Omega)$.
\end{definition}

The last term (in blue) in Eq.~\ref{eq:FEMNP}, and the second last term (in blue) in Eq.~\ref{eq:FEMNS} are the SUPG stabilization terms, while the last term in Eq.~\ref{eq:FEMNS} is the pressure stabilized petro-galerkin (PSPG) stabilizer. Where, $\tau_{SUPG}$ and $\tau_{PSPG}$ are element based standard coefficients for the SUPG and PSPG terms~\citep{article:TezMitRayShi92}. 

\begin{remark}
    As our discretization is restricts the basis functions to discrete counterparts of~$\vec{H}_0^1$~and ~${H}^1$ spaces.  Terms in the SUPG and PSPG residual which require higher regularity than $\vec{H}_0^1$ and ${H}^1$ are therefore neglected.  For example, the drift terms in the Nernst-Planck equations given by, 
    \begin{equation}
        \sum_{K=1}^{N_{el}}\Big(\tau_{SUPG}(\nabla\phi^{h} + \vec{u}^{h}) \cdot\nabla q_i^{h},
			\nabla \cdot (\nabla c^{h}_{i} + z_{i} c^{h}_{i}\nabla \phi^h)\Big) \,\, ,
    \end{equation}
    would require representation of second order derivatives on $c^{h}_i$ and $\phi^{h}$  which do not reside in~${H}^1$ space.
\end{remark}


\subsection{Dirichlet-to-Neumann transformation for the Poisson-Nernst-Plank equation}\label{section:weakBC}
Now, we present the formulation for weakly imposing Dirichlet boundary conditions.  
Without loss of generality, we consider the Dirichlet-to-Neumann transformation on the boundary, $\Gamma_D$. On this boundary, Dirichlet conditions are imposed on the species concentration, $c_i = g_{ci}$, and potential $\phi = g_{\phi}$. The Dirichlet-to-Neumann transformation replaces the strong imposition of these boundary conditions by a set of three boundary integral terms~\citep{bazilevs2007weak} --- representing the standard weakening (for consistency), its adjoint, and a penalty term. The penalty term ensures that as the mesh is refined, the strong imposition (i.e. Dirichlet condition) of boundary condition is recovered. 

The variational form of the NS--PNP equations including these three additional terms (in red) for the Poisson, and the Nernst-Plank equation is given as:
\begin{definition}
	find $\vec{u}(\vec{x}) \in \vec{H}_0^{1,h}(\Omega)$, $c_1(\vec{x}),...,c_N(\vec{x}), \phi(\vec{x}), p(\vec{x})$ $\in {H}^{1,h}(\Omega)$ such that
	\begin{align}
		\begin{split}
			\text{Nernst Planck Eqns:} & \quad
			\mathcal{B}_{NP,i}^{h}\Big(q_i^{h};c_i^{h},\phi^{h},\vec{u}^{h}\Big) 
			+ \mathcal{L}_{NP,i}^{h}\Big(q_i^{h};c_i^{h},\phi^{h}\Big) \\
			& \quad\quad\quad\quad\quad\quad\quad\quad\quad\quad 
			+ \textcolor{blue}{\sum_{K=1}^{N_{el}}\Big(\tau_{SUPG}(\nabla\phi^{h} + \vec{u}^{h}) \cdot\nabla q_i^{h},
			\frac{\partial c_i^h}{\partial t} + \vec{u}^h \cdot \nabla c_{i}^h\Big)} \\
			& \quad\quad\quad\quad\quad\quad\quad\quad\quad\quad
			- \textcolor{red}{\left(q_i^{h},\nabla c_i^{h} \cdot \vec{n} + z_{i}c_i^{h} \nabla \phi^{h} \cdot \vec{n}\right)_{\Gamma_D}}\\
			& \quad\quad\quad\quad\quad\quad\quad\quad\quad\quad
			- \textcolor{red}{\left(\nabla q_i^{h}\cdot\vec{n},c_i^{h} - g_{ci}\right)_{\Gamma_D}}\\
			& \quad\quad\quad\quad\quad\quad\quad\quad\quad\quad 
			+\textcolor{red}{\left(\frac{C_{NP}}{h_{el}}q_i^h,c_i^{h} - g_{ci}\right)_{\Gamma_D}} = 0, ~\text{for}~i = 1,...,N, 
		\end{split}\label{eq:FEMNP_wBC}\\
		\begin{split}
			\text{Poisson Eqn:}& \quad
			\mathcal{B}_{P}^{h}\Big(q^{h};\phi^{h}\Big) 
			+ \mathcal{L}_{P}^{h}\Big(q^{h};\phi^{h}\Big) 
			- \textcolor{red}{2\Lambda^2\left(q^h,\nabla \phi^h \cdot \vec{n}\right)_{\Gamma_D}}\\
			& \quad\quad\quad\quad\quad\quad\quad\quad\quad\quad\quad\quad
			- \textcolor{red}{2\Lambda^2\left(\nabla q^h\cdot \vec{n},\phi^h-g_{\phi}\right) _{\Gamma_D}}\\
			& \quad\quad\quad\quad\quad\quad\quad\quad\quad\quad\quad\quad
			+ \textcolor{red}{\left(\frac{C_P}{h_{el}}q_i^h,c_i^{h} - g_{\phi}\right)_{\Gamma_D}} = 0,
		\end{split}\label{eq:FEMPoisson_wBC}\\
		\begin{split}
			\text{Navier-Stokes Eqns:}& \quad
			\mathcal{B}_{NS}^{h}\Big(\vec{w}^{h},q^{h};\vec{u}^{h},p^{h}\Big) 
			+ \mathcal{L}_{NS}^{h}\Big(\vec{w}^{h},\vec{u}^{h}\Big)\\
			& \quad\quad\quad\quad\quad\quad\quad\quad\quad\quad
			+ \textcolor{blue}{\sum_{K=1}^{N_{el}}\Big(\tau_{SUPG}\vec{u}^{h}\cdot\nabla \vec{w}^{h},
				\frac{\partial \vec{u}^{h}}{\partial t} + \vec{u}^{h}\cdot \nabla \vec{u}^{h} - \vec{f_b}^{h}\Big)}\\
			& \quad\quad\quad\quad\quad\quad\quad\quad\quad\quad
			+ \textcolor{blue}{\sum_{K=1}^{N_{el}}\Big(\tau_{PSPG}\nabla{q}^{h},
				\frac{\partial \vec{u}^{h}}{\partial t} + \vec{u}^{h}\cdot \nabla \vec{u}^{h} - \vec{f_b}^{h} \Big)}= 0,
		\end{split}\label{eq:FEMNS_wBC}
	\end{align}
	$\forall \vec{w} \in \vec{H}^{1,h}_0(\Omega)$, $\forall q \in H^{1,h}(\Omega)$.
\end{definition}

The last term in Eq.~\ref{eq:FEMNP_wBC} and Eq.~\ref{eq:FEMPoisson_wBC} are the penalty-like terms \cite{bazilevs2007weak}. \(C_{NP}\) and \(C_{P}\) are the penalty coefficients that are specified based on inverse element estimates~\citep{article:TezMitRayShi92,Harari1992}. We set them equal to 4 for the simulation results shown in this work. The first two terms in red in Eq.~\ref{eq:FEMNP_wBC} and Eq.~\ref{eq:FEMPoisson_wBC} represent, respectively, the consistency term (arising from weakening the highest derivative terms in these equations), and the adjoint consistency term. The adjoint consistency term ensures better conditioning of the ensuing stiffness matrix.

\begin{remark}
    The weak imposition of Dirichlet boundary conditions on NS is not considered, as the main focus of the current study is the charged species transport in microfluidic applications. Low \(Re\) in microfluidic applications ensures reasonably large fluid boundary layers \footnote{We do not consider electroconvection~\cite{mani2020electroconvection}, where steep gradients in both concentration and velocity are expected. It is straightforward to incorporate weak boundary conditions for velocity~\cite{bazilevs2007weak}. We defer this exercise to later work.}; thus, the necessity of weakly imposed boundary condition for the Navier-Stokes equation diminishes. 
    \end{remark}
    \begin{remark}
    We use a block iterative approach for solving the Poisson-Nernst-Planck, and Navier-Stokes equations per time step. This ensures decoupled treatment of the \(c_i\) and \(\phi\) terms in the boundary terms. Block iteration between the Poisson and Nernst-Planck equations allows separate weak BC implementation for \(c_i\) and \(\phi\), as \(\phi\) can be treated like a constant during the iteration for NP. In addition, block iteration removes the non-linearity in the NP equation, see Section \ref{section:NumericalMethod} for more details on numerical methods.
    \end{remark} 
\section{Strategy for implementation} \label{section:NumericalMethod}
As specified before, we use a block iterative strategy to solve the set of equations. This approach provides several advantages, including (a) reducing the number of degrees of freedom per solve, (b) mitigating the numerical stiffness that exists between the equations (especially the large body force in the momentum equation), (c) enabling (simplified) weak imposition of Dirichlet boundary conditions by allowing separate treatment for \(c_i\) and \(\phi\) (see the boundary condition terms in Eq.~\ref{eq:FEMNP_wBC} and Eq.~\ref{eq:FEMPoisson_wBC}), and (d) converting the non-linear PNP equation into a set of two linear equations -- Poisson and Nernst-Planck. 

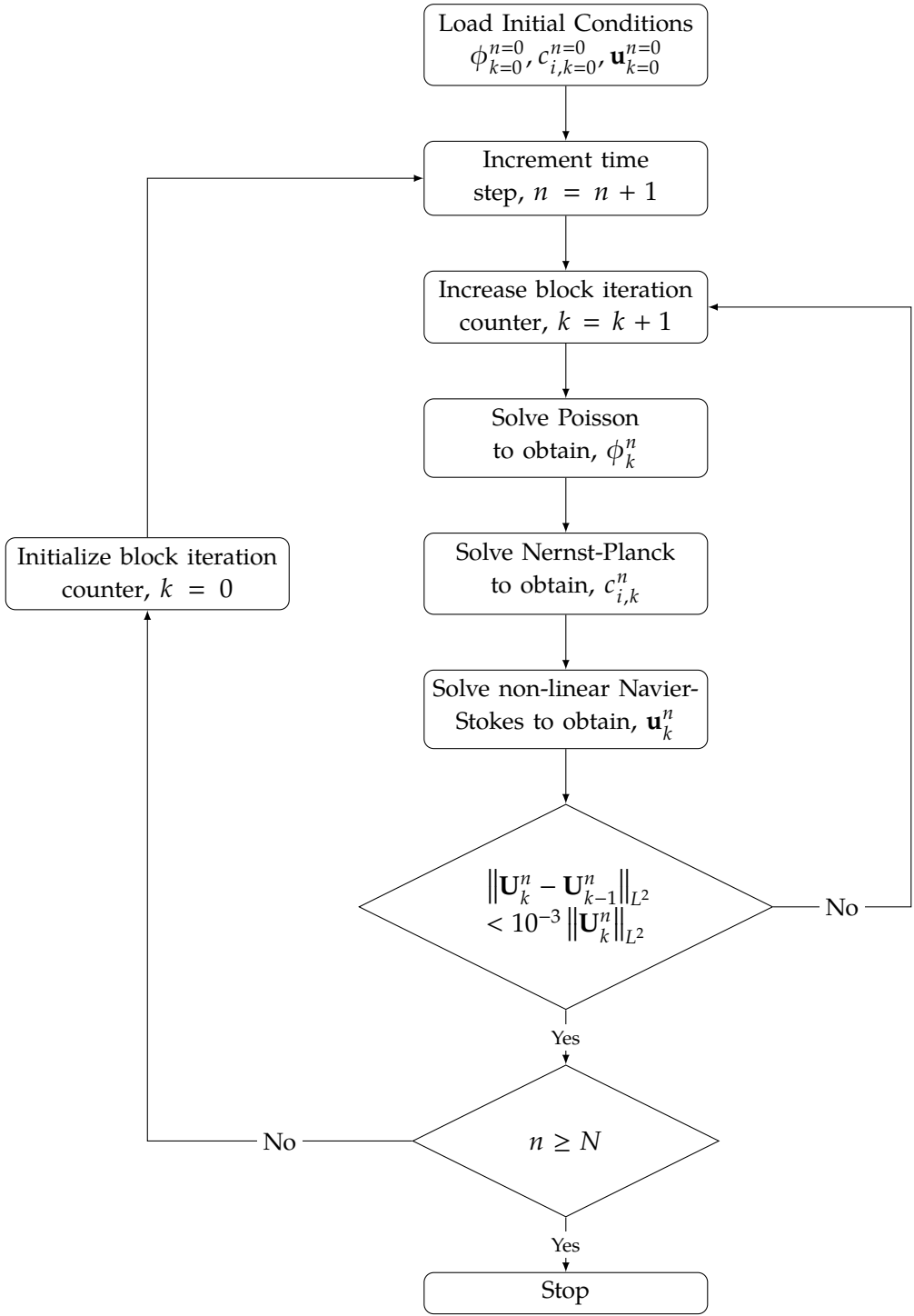
\begin{figure}[!hbtp]
\begin{tikzpicture}
    \matrix (m)[matrix of nodes, column  sep=1cm,row  sep=8mm, align=center, nodes={rectangle,draw, anchor=center} ]{
    &    |[block]| {{\small Load Initial Conditions} \(\phi_{k=0}^{n=0}, c_{i,k=0}^{n=0}, \vec{u}_{k=0}^{n=0}\)} &                \\
    &    |[block]| {{\small Increment time step}, $n=n+1$}  &                \\
    &    |[block]| {{\small Increase block iteration counter}, $k=k+1$}    &                \\
    &    |[block]| {{\small Solve Poisson to obtain}, \(\phi_{k}^{n}\)}   &                \\
|[block]| {{\small Initialize block iteration counter}, $k=0$}    &    |[block]| {{\small Solve Nernst-Planck to obtain}, \(c_{i,k}^{n}\)}  &            \\
    &    |[block]| {{\small Solve non-linear Navier-Stokes to obtain}, \(\vec{u}_{k}^{n}\)}   &   \\
    &    |[decision]| {
    	$\norm{\vec{U}_{k}^{n}-\vec{U}_{k-1}^{n}}_{L^2}$ $<\SI{e-3}{}\norm{\vec{U}_{k}^{n}}_{L^2}$}                                 &                \\
    &    |[decision]| {\(n \geq N\)}                                 &                \\
    &    |[block]| {{\small Stop}}   &   \\
    };
    \path [>=latex,->] (m-1-2) edge (m-2-2);
    \path [>=latex,->] (m-2-2) edge (m-3-2);
    \path [>=latex,->] (m-3-2) edge (m-4-2);
    \path [>=latex,->] (m-4-2) edge (m-5-2);
    \path [>=latex,->] (m-5-2) edge (m-6-2);
    \path [>=latex,->] (m-6-2) edge (m-7-2);
    \draw [rectangle connector=5cm] (m-7-2) to node[descr] {{\small No}} (m-3-2);
    \draw [straight connector] (m-7-2) to node[descr] {Yes} (m-8-2);
    \draw [straight connector] (m-8-2) to node[descr] {Yes} (m-9-2);
    \draw [>=latex,->] (m-8-2) -| node[descr, pos=0.25] {{\small No}} (m-5-1);
    \draw [>=latex,->] (m-5-1) |- (m-2-2);
\end{tikzpicture}
\caption{Flow chart of NS-PNP solver} \label{fig:Flowchart}
\end{figure}

A flow chart of the approach is illustrated in Figure ~\ref{fig:Flowchart}. We utilize a Backward Euler time step for all equations. We implement a parallel version of this method within our in-house parallel finite element framework. The domain decomposition is achieved via ParMETIS~\cite{karypis1997parmetis}. 
We make use of the  {\textsc{Petsc}} library, which provides efficient parallel implementations of linear and non-linear solvers along with an extensive suite of preconditioners~\citep{petsc-efficient, petsc-web-page, petsc-user-ref}. Specifically, we utilize the SNES construct (line search quasi-Newton) for the Navier-Stokes solver, and the KSP construct for the linear system. 

\section{Numerical experiments}\label{sec:results}


\subsection{Convergence against manufactured solution}
\begin{figure}
    \centering
    \begin{tikzpicture}
    \tikzstyle{every node}=[font=\footnotesize]
      \begin{loglogaxis}[
          width=0.7\linewidth, 
          grid=both,
          grid style={black!50,dotted},
          xmode=log, ymode=log,
          xlabel=$1/h$, 
          ylabel=$\|e\|_{L^2(\Omega)}$,
          legend style={at={(0.95,0.95)},anchor=north east},
          x tick label style={rotate=0,anchor=north}, 
          axis line style = thick,
          cycle list name=color list
        ]
        \addplot[black, dash dot, mark = o, mark size=2pt, mark options={solid}, line width=0.4mm] 
        table[x expr={1.0/((\thisrow{h}))},y expr={\thisrow{u}},col sep=space]{5.RESULTS/data/L2.txt};
        
        \addplot[orange, loosely dotted, mark = +, mark size=2pt, mark options={solid}, line width=0.4mm]
        table[x expr={1.0/((\thisrow{h}))},y expr={\thisrow{v}},col sep=space]{5.RESULTS/data/L2.txt};
        
        \addplot[teal, dashed, mark = x, mark size=2pt, mark options={solid}, line width=0.4mm]
        table[x expr={1.0/((\thisrow{h}))},y expr={\thisrow{cp}},col sep=space]{5.RESULTS/data/L2.txt};
        
        \addplot[blue, densely dotted, mark = diamond, mark size=3pt, mark options={solid}, fill = none, line width=0.4mm]
        table[x expr={1.0/((\thisrow{h}))},y expr={\thisrow{cn}},col sep=space]{5.RESULTS/data/L2.txt};
        
        \addplot[olive, dash dot dot, mark = square, mark size=2pt, mark options={solid}, line width=0.4mm]
        table[x expr={1.0/((\thisrow{h}))},y expr={\thisrow{phi}},col sep=space]{5.RESULTS/data/L2.txt}
        coordinate [pos=0.3] (A) 
        coordinate [pos=0.45] (B);
        \draw (A) |- (B)  
        node [pos=0.3, anchor=east] {2} 
        node [pos=0.75, anchor=north] {1}; 
        \legend{$u$,$v$,$c_+$,$c_-$,$\phi$}
    \end{loglogaxis}
\end{tikzpicture}
    \caption{Spatial convergence of the NS-PNP solver using method of manufactured solution.}
    \label{fig:mms}
\end{figure}
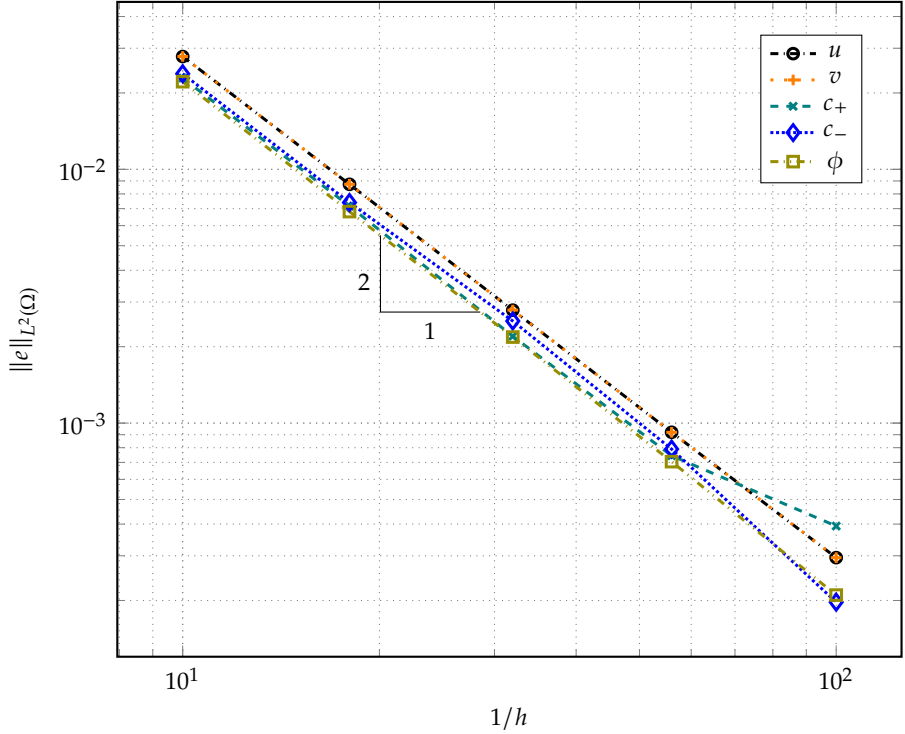

We use the method of manufactured solutions to assess the convergence of our implementation.  We select an input ``solution'', and substitute it in the full set of governing equations. We then use the residual as a body force on the right-hand side of Eqs~\ref{eq:FEMNP},\ref{FEMPoisson},\ref{eq:FEMNS}.
We choose the following ``solution'' with appropriate body forcing terms: 
\begin{equation}\label{eq:mms}
\begin{cases}
    u = cos(2\pi t)sin(2\pi x)cos(2\pi y), \\
    v = -cos(2\pi t)cos(2\pi x)sin(2\pi y), \\
    p = cos(2\pi t)sin(2\pi x)cos(2\pi y),\\
    c_+ = cos(2\pi t)cos(2\pi x)sin(2\pi y),\\
    c_- = cos(2\pi t)sin(2\pi x)cos(2\pi y),\\
    \phi = -cos(2\pi t)cos(2\pi x)sin(2\pi y).
\end{cases}    
\end{equation}
Note that the manufactured solution for the fluid velocity is divergence free. Our numerical experiments use the following non-dimensional parameters: $\Lambda = 10^{-2}$, $Sc = 1$, $\kappa = 1.0$. We fix the time step at $k = 10^{-4}$ to minimize contribution of error from temporal discretization.  We vary the spatial mesh resolution by increasing the number of elements. Figure \(\ref{fig:mms}\) shows the spatial convergence of $L^2$ errors (numerical solution compared with the manufactured solution) at $t = 1$. We observe second order convergence for velocity, species concentration and potential as expected for linear conforming Galerkin basis functions.


\subsection{Electroosmotic flow (EOF) simulation and comparison with analytical results}

Electroosmotic flow is a canonical microfluidic flow where the flow is driven by a potential drop \(\Delta \phi\) maintained across a channel with charged walls~\cite{kirby2010micro}. In the bulk solution away from the charged wall, charge neutrality is maintained (\(\rho_e = 0\)). Hence, the bulk fluid does not respond to the applied potential drop. However, the charged wall attracts counter-ions and expels co-ions breaking the charge neutrality (\(\rho_e\neq 0\)) in the fluid domain that is wall adjacent. This results in a non-zero body force term in the Navier-Stokes equation near the walls. Subsequently, the rest of bulk fluid is driven to a steady state profile by the shear stress from the near wall flow. This flow profile has a characteristic plug shape, which is distinct from pressure driven Poiseuille flow in microchannels. The plug velocity can be analytically computed and is given by~\cite{kirby2010micro}
\begin{equation}\label{eq:EOFAnal}
    U_{max} = -\frac{\varepsilon \phi_0}{\eta}E.
\end{equation}
where \(\phi_0\) is potential difference between the channel wall and bulk fluid.

\begin{figure}[!hbtp]
\centering
\includegraphics[trim={0 5cm 0 2.5cm},clip,width=\textwidth]{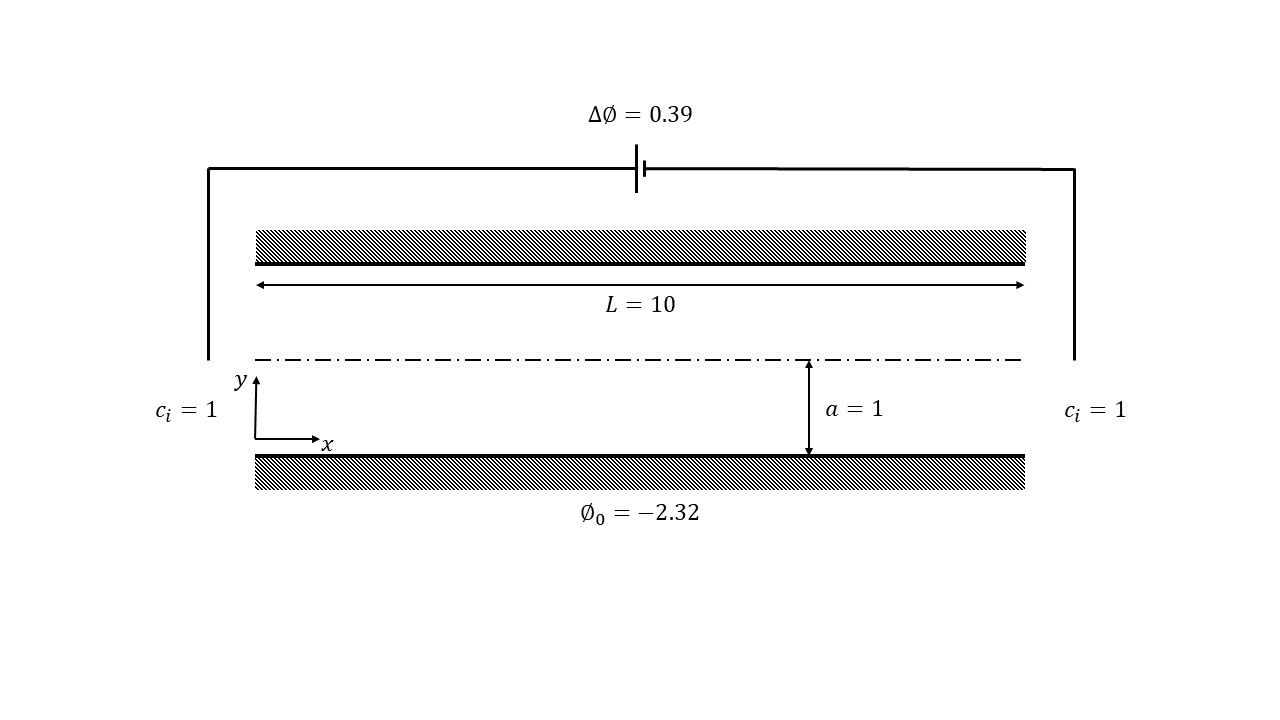}
    \caption{Problem geometry and boundary conditions for EOF simulation.}
\label{fig:EOF}
\end{figure}

We demonstrate the ability of the NS-PNP framework to accurately predict this EOF profile. The boundary conditions and problem geometry are shown in Figure \(\ref{fig:EOF}\). The dimensional values of all quantities are provided in Table \ref{Tab:EOF}, while all simulations are performed in dimensionless terms. The applied potential difference per unit length across the channel was $\Delta \Phi = 0.039$, the wall potential was $\Phi_0= -2.32$, and the inlet and outlet cation \(c_+\) and anion \(c_-\) concentrations are set to 1. The charge valences of the species, \(z_i\), are 1 and -1, respectively. The dimensionless Debye layer thickness, \(\Lambda\) was 0.097, Schmidt number, \(Sc\) was 686.68, and electrohydrodynamic coupling constant was 0.4037. The analytic solution for $U_{max}$ from Eq.~\ref{eq:EOFAnal} gives a non-dimensional value of 0.0429 (and dimensional value of $\SI{5.5711e-04}{m/s}$). 


\begin{table}
\centering
\begin{tabular}{c|cc} 
\toprule
\multicolumn{1}{l}{} & Non-dimensional & Dimensional  \\ 
\hline\hline
\(\Lambda\)  &  0.0971      &  N/A                                      \\
\(\lambda\)  &  N/A         &  \SI{9.7085}{\newton\meter}               \\
\(L\)        &  N/A         &  \SI{1e-7}{\meter}                        \\
\(E\)        &  0.0387      &  \SI{10000}{\volt\per\meter}              \\
\(\phi_0\)   &  -2.3202     &  \SI{-0.07}{\volt}                        \\
\(\epsilon\) &  N/A         &  \SI{7.0832e-10}{\farad\per\meter}        \\
\(Sc\)       &  686.6754    &  N/A                                      \\
\(\mu\)      &  N/A         &  \SI{0.89}{\milli\pascal\second}          \\
\(D\)        &  N/A         &  \SI{1.3e-9}{\meter\squared\per\second}   \\
\(\rho\)     &  N/A         &  \SI{997}{\kilo\gram\per\meter\cubed}     \\
\bottomrule
\end{tabular}
\caption{Dimensional and non-dimensional properties for EOF simulation.}
\label{Tab:EOF}
\end{table}

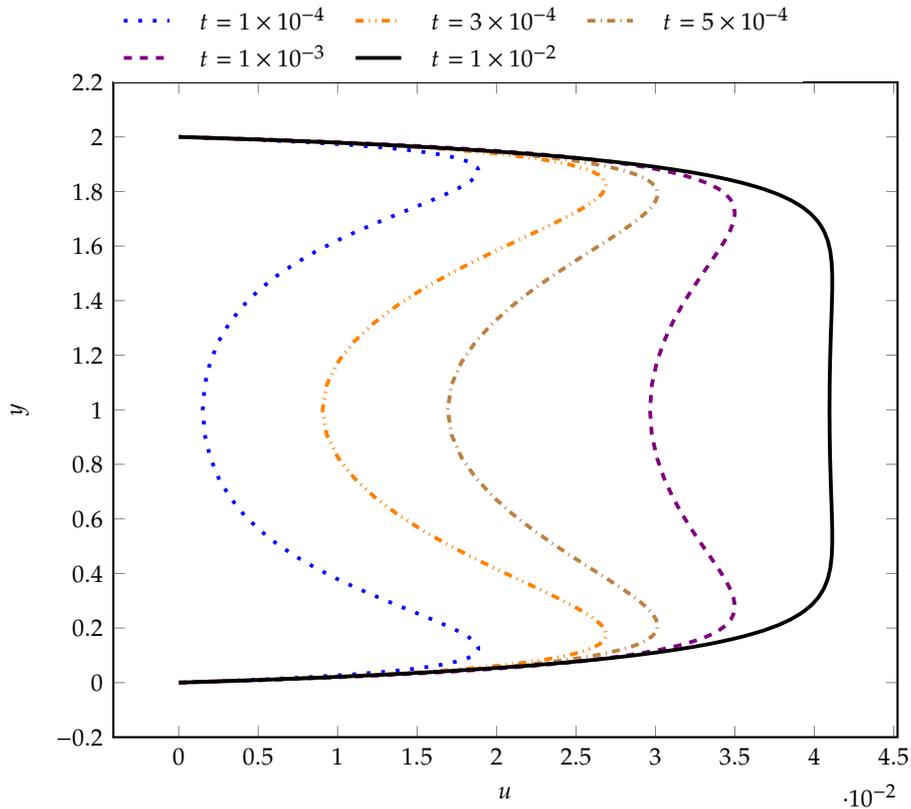
\begin{figure}
    \centering
    \begin{tikzpicture}
    \tikzstyle{every node}=[font=\footnotesize]
      \begin{axis}[
          width=0.7\linewidth, 
          xlabel=$u$, 
          ylabel=$y$,
          legend style={at={(0,1)},anchor=south west,legend columns=3, column sep=0.3cm, draw=none}, 
          x tick label style={rotate=0,anchor=north}, 
          axis line style = thick,
          cycle list name=color list
        ]
        \addplot[blue, loosely dotted, line width=0.5mm]
        table[x expr={\thisrow{UX1}},y expr={\thisrow{Y}},col sep=space]{5.RESULTS/data/eofData.txt};
        
        \addplot[orange, dash dot dot, line width=0.5mm]
        table[x expr={\thisrow{UX2}},y expr={\thisrow{Y}},col sep=space]{5.RESULTS/data/eofData.txt};
        
        \addplot[brown, dash dot, line width=0.5mm]
        table[x expr={\thisrow{UX3}},y expr={\thisrow{Y}},col sep=space]{5.RESULTS/data/eofData.txt};
        
        \addplot[violet, dashed, line width=0.5mm]
        table[x expr={\thisrow{UX4}},y expr={\thisrow{Y}},col sep=space]{5.RESULTS/data/eofData.txt};
        
        \addplot[solid, line width=0.5mm]
        table[x expr={\thisrow{UX5}},y expr={\thisrow{Y}},col sep=space]{5.RESULTS/data/eofData.txt};
        
        \legend{$t=\SI{1e-4}{}$,$t=\SI{3e-4}{}$,$t=\SI{5e-4}{}$,$t=\SI{1e-3}{}$,$t=\SI{1e-2}{}$}
    \end{axis}
\end{tikzpicture}
    \caption{Time evolution of EOF velocity profile in non-dimensional time.}
    \label{fig:EOFProfile}
\end{figure}

We discretize the domain into $200 \times 40$ linear elements and use a time step of $k = 10^{-4}$. Figure \ref{fig:EOFProfile} shows the time evolution of the velocity profile. At early times, the body force results in non-zero fluid velocity only in the wall adjacent regions. This near wall flow subsequently drives the bulk. After about 100 time steps, the flow profile nearly reaches steady state, exhibiting the classic plug shape. As seen from Figure \ref{fig:EOFProfile}, the computed value of $U_{max}$ is $0.0413$. 

\subsection{Electrokinetics near a permselective membrane: 1D simulations and flux comparisons} \label{1D_IDZ}
We next illustrate the framework for practical application involving electrokinetics near permselective membranes, which is an area of research that is seeing increasing interest. In particular, we showcase how the weak enforcement of boundary conditions allows accurate capture of current fluxes at boundaries without very fine mesh resolution. A permselective membrane selectively transports species forming a depletion zone and an enrichment zone at the opposite sides of the membrane \cite{li2016recent}. For example, Nafion is a cation selective membrane that is widely used in electrokinetic applications. Under an applied electric current, Nafion selectively transfers cations across the membrane, while blocking anions. This behavior is critical for a wide variety of applications including separation of biological entities \cite{berzina2018electrokinetic} and sea water desalination \cite{kim2010direct}.


Our model accurately predicts the formation of a depletion zone near the permselective membrane. A simple binary electrolyte ($N = 2$, \(z_1\) = 1 and \(z_2\)= -1) was considered for the simulation. The cation selective membrane was located at \(x=0\), and the bulk electrolyte is at \(x=1\). \(\phi(x=0) =\) 0 at the membrane, and \(\phi(x=1) =\) 50 at the bulk. Both \(c_+\) and \(c_-\) were set to 1 at \(x=1\). The non-dimensional Debye length, \(\Lambda\), was 0.01. \(c_+\) was set to 2 at the membrane. 

\begin{figure}
    \centering
    \begin{tikzpicture}
    \tikzstyle{every node}=[font=\footnotesize]
      \begin{axis}[
          width=0.7\linewidth, 
          xlabel=$x$, 
          legend style={at={(0.95,0.95)},anchor=north east}, 
          legend cell align={left},
          x tick label style={rotate=0,anchor=north}, 
          axis line style = thick,
          cycle list name=color list
        ]
        \addplot[black, solid, line width=0.5mm]
        table[x expr={\thisrow{x}},y expr={\thisrow{phi}/50},col sep=space]{5.RESULTS/data/depStr.txt};
        
        \addplot[blue, dashed, line width=0.5mm]
        table[x expr={\thisrow{x}},y expr={\thisrow{cp}},col sep=space]{5.RESULTS/data/depStr.txt};
        
        \addplot[red, dash dot dot, line width=0.5mm]
        table[x expr={\thisrow{x}},y expr={\thisrow{cn}},col sep=space]{5.RESULTS/data/depStr.txt};
        
        \legend{$0.02\times\phi$,$c_+$,$c_-$}
    \end{axis}
\end{tikzpicture}
    \caption{Species concentrations and potential near the cation selective membrane (\(x=0\)). The bulk is at \(x=1\).}
    \label{fig:depStr}
\end{figure}
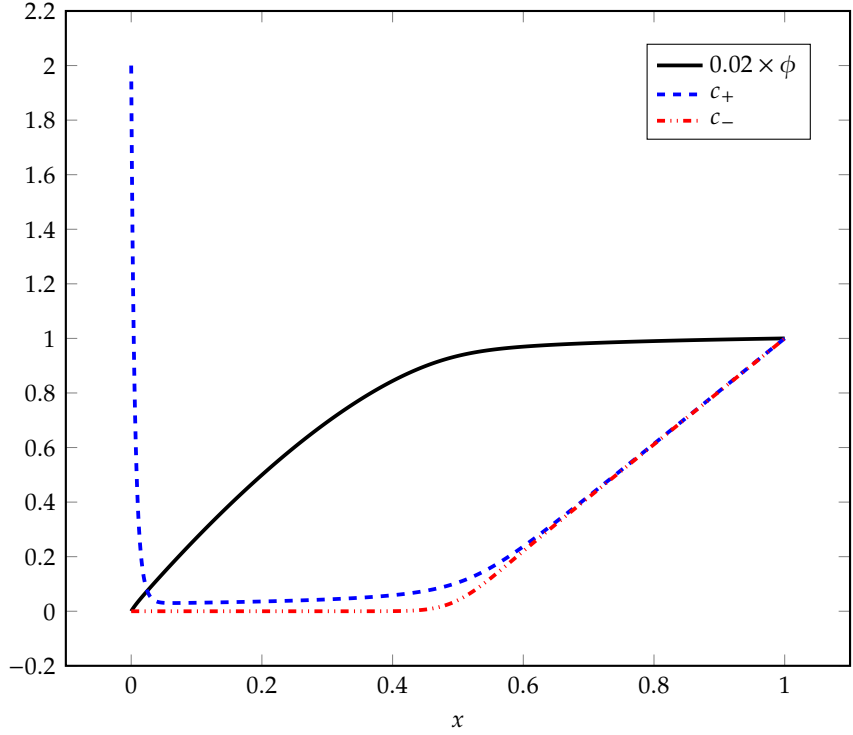

The results of the PNP calculation with strongly imposed boundary conditions are shown in Figure \ref{fig:depStr} as the baseline. As explained above, both cations and anions were depleted near the cation selective membrane. The magnitude of the electric field (absolute value of the potential gradient) was high in the depletion zone and drops as it extends into the bulk. Adjacent to the membrane, a thin boundary layer of cations forms. 
The thickness of the thin boundary layer is proportional to the non-dimensional Debye length, \(\Lambda\). Thus, a very thin concentration boundary layer is expected. Traditionally, without significant mesh resolution, inaccurate evaluation of the stiff gradients at the boundary result in significant error in the current flux calculation. Accurate evaluation of charge flux is especially critical, because in most electrokinetic or electrochemical experiments, flux is the single most important measurement used to understand the system \cite{bard2000electrochemical}. 

The weak imposition of Dirichlet boundary conditions allows relaxation of the mesh resolution requirements, while retaining accuracy of boundary flux computations. The boundary flux at the membrane was calculated considering global conservation \cite{bazilevs2007weak}, and setting the test basis function to 1, resulting in: 
\begin{equation}\label{eq:flxCal}
  \textit{Species i flux,}~~\vec{j}_i \cdot \vec{n} = - \left(1,\nabla c_i^{h} \cdot \vec{n} + z_{i} c_i^{h} \nabla \phi^{h} \cdot \vec{n}\right)_{\Gamma_D}	+\left(\frac{C_{NP}}{h_{el}}1,c_i^{h} - g_{ci}\right)_{\Gamma_D} 
\end{equation} 

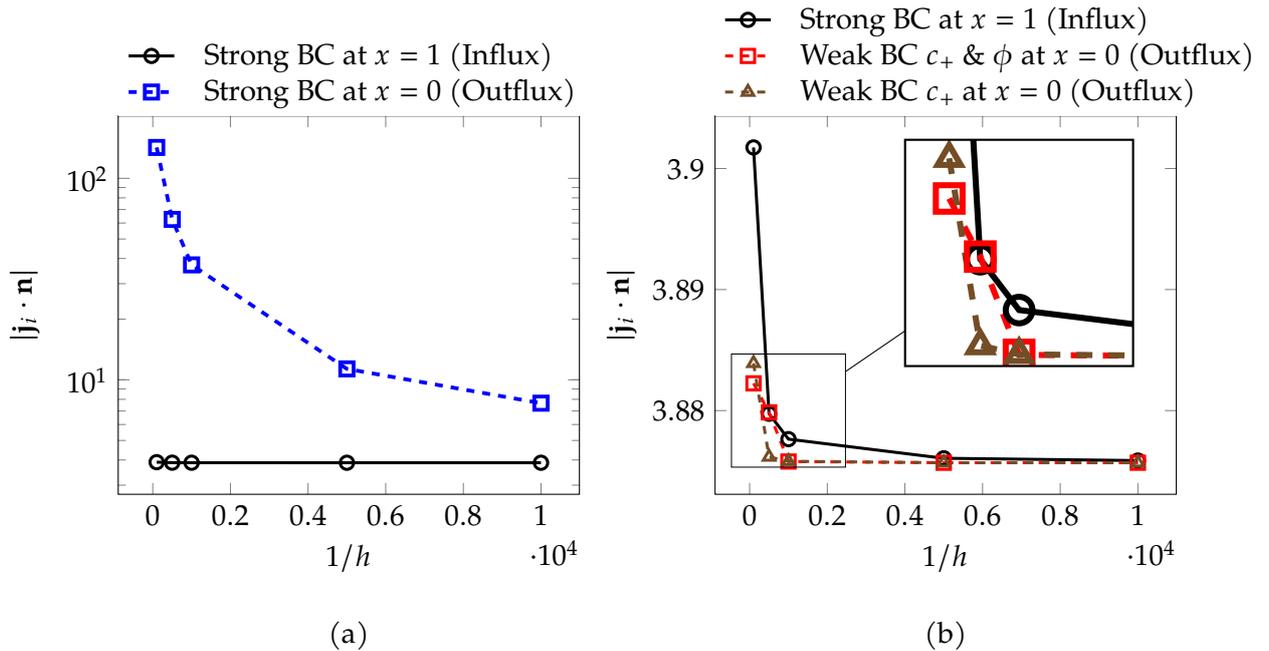
\begin{figure}[b!]
\centering
\begin{tikzpicture}[]
  \begin{semilogyaxis}[
    width=0.45\linewidth, 
    title = {(a)},
    title style={at={(0.5,0)},anchor=north,yshift=-50},
    xlabel= $1/h$,
    ylabel= \(|\vec{j}_{i}\cdot\vec{n}|\),
    legend style={at={(0,1)},anchor=south west,column sep=0.3cm, draw=none},
    legend cell align={left}
  ]
        \addplot+[black, 
            mark = o, 
            mark size=2.5pt, mark options={solid}, line width=0.4mm]
            table[x expr={\thisrow{inv_h}},y
            expr={\thisrow{strIn}},col sep=space]{5.RESULTS/data/1Dflx.txt};
        \addplot+[blue, dashed, 
                mark = square, 
                mark size=2.5pt, mark options={solid}, line width=0.5mm]
            table[x expr={\thisrow{inv_h}},y expr={\thisrow{strOut}},col sep=space]{5.RESULTS/data/1Dflx.txt}; 
        \legend{{Strong BC at $x = 1$ (Influx)}, {Strong BC at $x=0$ (Outflux) }}
\end{semilogyaxis}
\end{tikzpicture}
\begin{tikzpicture}[spy using outlines={rectangle, magnification=2, size=3cm, connect spies}]
  \begin{axis}[
    width=0.45\linewidth, 
    title = {(b)},
    title style={at={(0.5,0)},anchor=north,yshift=-50},
    xlabel= $1/h$,
    ylabel= \(|\vec{j}_{i}\cdot\vec{n}|\),
    legend style={at={(0,1)},anchor=south west,column sep=0.3cm, draw=none},
    legend cell align={left}
  ]
        \addplot+[black, 
                mark = o, 
                mark size=2.5pt, mark options={solid}, line width=0.4mm]
            table[x expr={\thisrow{inv_h}},y
            expr={\thisrow{strIn}},col sep=space]{5.RESULTS/data/1Dflx.txt};
        \addplot+[
                dashed, 
                mark = square, 
                mark size=2.5pt, mark options={solid}, line width=0.4mm]
            table[x expr={\thisrow{inv_h}},y expr={\thisrow{weakPC}},col sep=space]{5.RESULTS/data/1Dflx.txt}; 
        \addplot+[
                dashed, 
                mark = triangle, 
                mark size=2.5pt, mark options={solid}, line width=0.4mm]
            table[x expr={\thisrow{inv_h}},y expr={\thisrow{weakC}},col sep=space]{5.RESULTS/data/1Dflx.txt};
        \legend{{Strong BC at $x=1$ (Influx)}, {Weak BC $c_+$ \& $\phi$ at $x=0$ (Outflux)},{Weak BC $c_+$ at $x=0$ (Outflux)}}
        \coordinate (a) at (axis cs:1000,3.88);
\end{axis}
\spy [black] on (a) in node  at (4,3.2);
\end{tikzpicture}
\caption{Boundary flux calculation at the membrane ($x=0$) using strong (panel (a)) and and weak (panel (b)) boundary conditions. On both figures, the (easy to compute) flux in the bulk ($x=1$) is plotted in black. Notice the large difference in the magnitude of the y-axis between the strong and weak imposition cases.}
\label{fig:1Dflx}
\end{figure} 

We compare the boundary flux calculation at different mesh resolutions between a strong imposition versus two types of weak imposition of the Dirichlet boundary conditions: \textit{Type 1:} weak BC for both Poisson and Nernst-Planck equations; \textit{Type 2:} weak BC only for Nernst-Planck equation. Note that only the first term in equation (\ref{eq:flxCal}) is used to compute the strong flux (which is equivalent to surface integral of equation (\ref{eq:NonDimFlx})), while both terms in equation (\ref{eq:flxCal}) are used to compute boundary fluxes under weak imposition. 

As the membrane only transports cations, and the applied potential is high at the bulk (\(x=1\)) and low at the membrane (\(x=0\)), the direction of cation flux is from the bulk to the membrane. Under steady state conditions, the influx of the cations from the bulk must be equal to the outflux at the membrane. Thus, the influx from the bulk provides a baseline to compare the outflux computed at the membrane using the three different approaches. As seen from Figure ~\ref{fig:depStr}, the gradients of the species concentrations and potential at the bulk ($x = 1$) are significantly smaller, hence we expect the flux computed from imposition of strong boundary conditions to provide accurate values here. 

The flux calculation results with various mesh sizes are shown in Figure~\ref{fig:1Dflx}. Notice that in panel (a) of Figure~\ref{fig:1Dflx} the outflux at the membrane (x=0) from strong imposition has still not converged to the influx from the bulk (even at fine mesh resolutions), the flux from weak imposition (panel (b) of Figure~\ref{fig:1Dflx}) has converged to the influx current even for dramatically coarse mesh sizes. For a range of mesh resolutions, we can see only a minute difference between the influx and the outflux calculated using weak BC. The enlarged plot in panel (b) of Figure~\ref{fig:1Dflx} shows that the difference between the fluxes when weak boundary conditions are imposed for both $\phi$ and $c_i$ vs only for $c_i$ is negligible. 

\begin{figure}
    \centering
    \begin{tikzpicture}[spy using outlines={rectangle, magnification=2.4, size=3.5cm, connect spies}]
    \tikzstyle{every node}=[font=\footnotesize]
      \begin{axis}[
          width=0.7\linewidth, 
          xlabel=$x$, 
          ylabel=$c_+$ or $c_-$,
          xmin = 0.0,
          ymin = 0.0,
          legend style={at={(0.95,0.95)},anchor=north east}, 
          legend cell align={left},
          x tick label style={rotate=0,anchor=north}, 
          axis line style = thick,
          cycle list name=color list
        ]
        \addplot[black, solid, line width=0.4mm]
        table[x expr={\thisrow{x}},y expr={\thisrow{cp}},col sep=space]{5.RESULTS/data/LCtrl/1e-2.txt};
        
        \addplot[black, dashed, line width=0.4mm]
        table[x expr={\thisrow{x}},y expr={\thisrow{cn}},col sep=space]{5.RESULTS/data/LCtrl/1e-2.txt};
        
        \addplot[blue, solid, line width=0.4mm]
        table[x expr={\thisrow{x}},y expr={\thisrow{cp}},col sep=space]{5.RESULTS/data/LCtrl/05e-2.txt};
        
        \addplot[blue, dashed, line width=0.4mm]
        table[x expr={\thisrow{x}},y expr={\thisrow{cn}},col sep=space]{5.RESULTS/data/LCtrl/05e-2.txt};
        
        \addplot[red, solid, line width=0.4mm]
        table[x expr={\thisrow{x}},y expr={\thisrow{cp}},col sep=space]{5.RESULTS/data/LCtrl/1e-3.txt};
        
        \addplot[red, dashed, line width=0.4mm]
        table[x expr={\thisrow{x}},y expr={\thisrow{cn}},col sep=space]{5.RESULTS/data/LCtrl/1e-3.txt};
        
        \legend{$c_{+} \ \Lambda = \num{1e-2}$, 
        $c_{-} \ \Lambda = \num{1e-2}$,
        $c_{+} \ \Lambda = \num{5e-3}$,
        $c_{-} \ \Lambda = \num{5e-3}$,
        $c_{+} \ \Lambda = \num{1e-3}$,
        $c_{-} \ \Lambda = \num{1e-3}$}
        
        \coordinate (spypoint) at (axis cs:0.07,0.07);
    \end{axis}
    \spy [black] on (spypoint) in node  at (4,5);
\end{tikzpicture}
    \caption{Species concentrations and potential near the cation selective membrane (\(x=0\)) with various non-dimensional Debye layer, \(\Lambda\).}
    \label{fig:depStrLCtrl}
\end{figure}
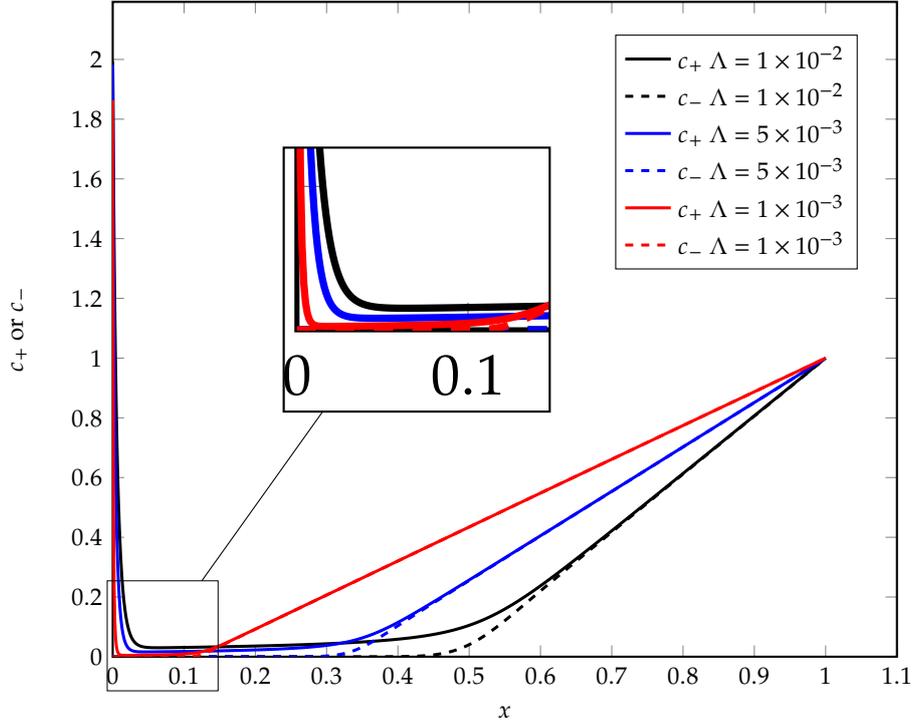

We next investigate electrokinetics near the membrane with various \(\Lambda\) (Figure~\ref{fig:depStrLCtrl} and Table~\ref{Tab:EOF_bFlx}) spanning two orders of magnitude. For all \(\Lambda\), the size of the mesh was set to \(h=\SI{1e-3}{}\). The boundary conditions were the same as the ones shown in Figure~\ref{fig:depStr}. Weak boundary conditions are applied at $x = 0$, while strong boundary conditions are applied at $x = 1$. Representative cation and anion distributions after steady state is reached are plotted in Figure~\ref{fig:depStrLCtrl}. As expected, with smaller \(\Lambda\), the thickness of the boundary layer decreases. We also see that the size of the depletion zone is correlated with \(\Lambda\). We compare the flux at $x = 0$ with the flux at $x = 1$ in Table \ref{Tab:EOF_bFlx}. As stated before, these fluxes should match at steady state and serve as an excellent validation test of the weakly imposed boundary condition. Across two orders in magnitude variation in \(\Lambda\), the fluxes reliably match, with a maximum deviation of less than $3\%$, even for the case when a single element is larger than the boundary layer (for $\Lambda = 5\times 10^{-4}$). We note that the calculated boundary flux decreased with decreasing \(\Lambda\), which agrees with other literature \cite{chu2005electrochemical}. 

\begin{table}
\centering
\begin{tabular}{c|cc} 
\toprule
\(\Lambda\)  &      flux (x = 0) weak BC         &        flux (x = 1) strong BC     \\
\hline\hline
\SI{5e-2}{}  &      \SI{16.47}{}       &        \SI{16.49}{}   
\\
\SI{1e-2}{}  &      \SI{3.88}{}       &        \SI{3.88}{}    \\
\SI{5e-3}{}  &      \SI{2.98}{}       &        \SI{2.98}{}    \\
\SI{1e-3}{}  &      \SI{2.28}{}       &        \SI{2.27}{}    \\
\SI{5e-4}{}  &      \SI{2.10}{}       &        \SI{2.16}{}    \\
\bottomrule
\end{tabular}
\caption{Weak outflux comparison with strong influx for various boundary layer thickness (\(\Lambda\))}
\label{Tab:EOF_bFlx}
\end{table}

\subsection{Electrokinetics near cation selective membrane: 2D simulations} 
\begin{figure}[!hbtp]
\centering
\begin{subfigure}{.45\textwidth}
    \centering
    \includegraphics[width = \linewidth]{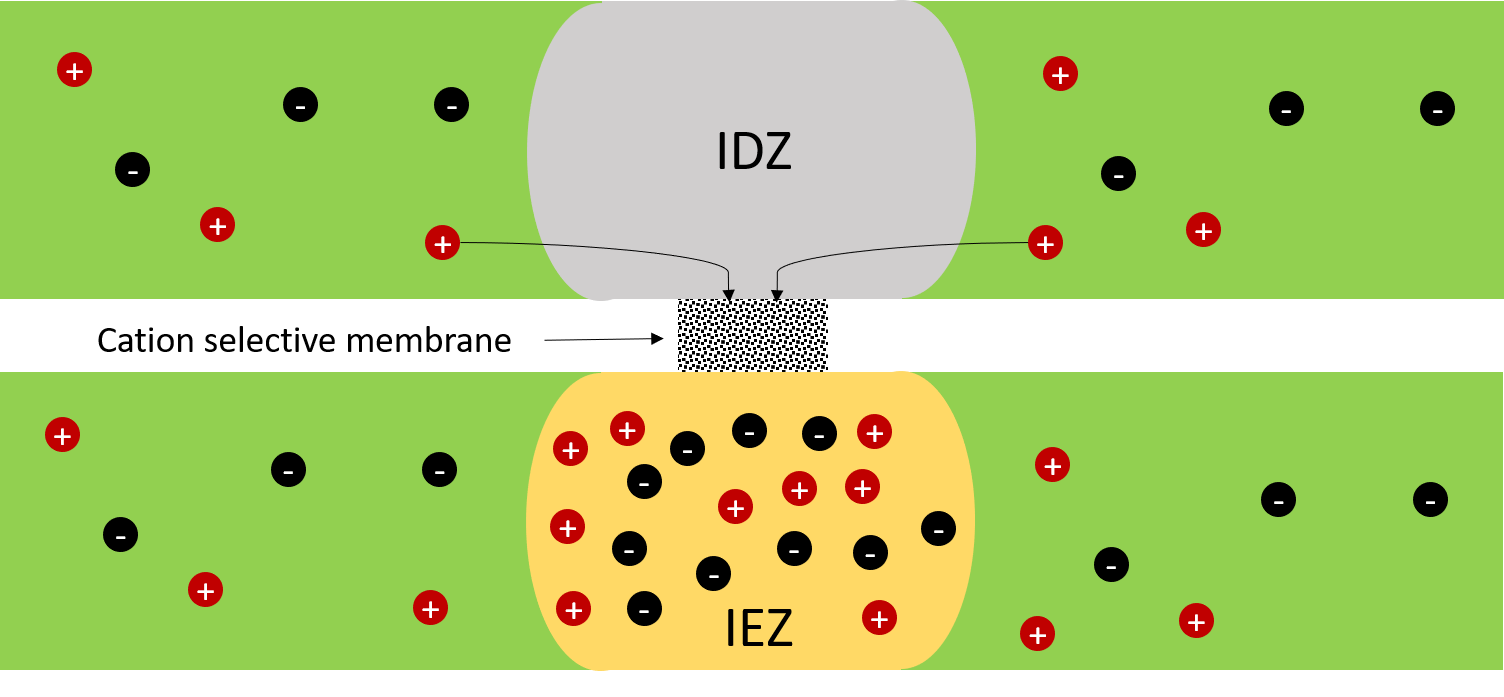}
    \subcaption{}
\end{subfigure}
\begin{subfigure}{.45\textwidth}
    \centering
    \includegraphics[width = \linewidth]{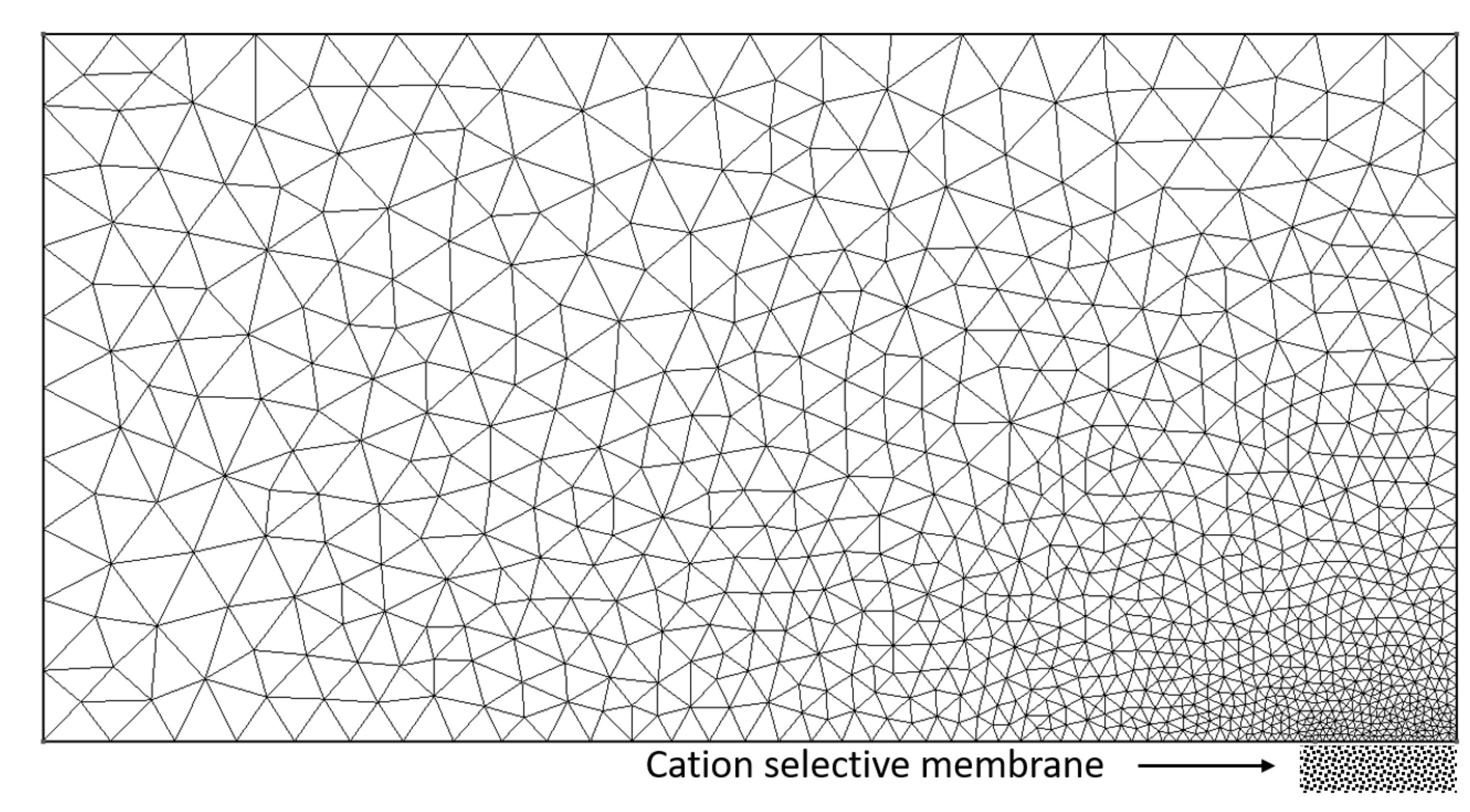}
    \subcaption{}
\end{subfigure}
\caption{Formation of IDZ and IEZ near the cation selective membrane (a), and the domain and meshes for the simulation (b). The left half of the IDZ channel was considered for the simulation.}
\label{fig:2DConcept}
\end{figure}

In this section, we illustrate the use of this approach to generate a 2D model of electrokinetic enrichment of a charged species near an IDZ generated by ICP in a microfluidic device. The device consists of two straight microchannels interconnected by a cation selective membrane (see Figure~\ref{fig:2DConcept}). A voltage bias is applied across the device through electrodes immersed in the fluid filled reservoirs of the two channels. The current resulting from the applied voltage is carried by cations and anions along the channels. The membrane transports only cations, while blocking anions, which creates an IDZ in the anodic channel and an IEZ in the cathodic channel~\cite{li2016recent}. 

\begin{figure}[!hbtp]
\centering
\includegraphics[width = 0.8\linewidth,trim=3cm 7.5cm 3cm 7.5cm,clip]{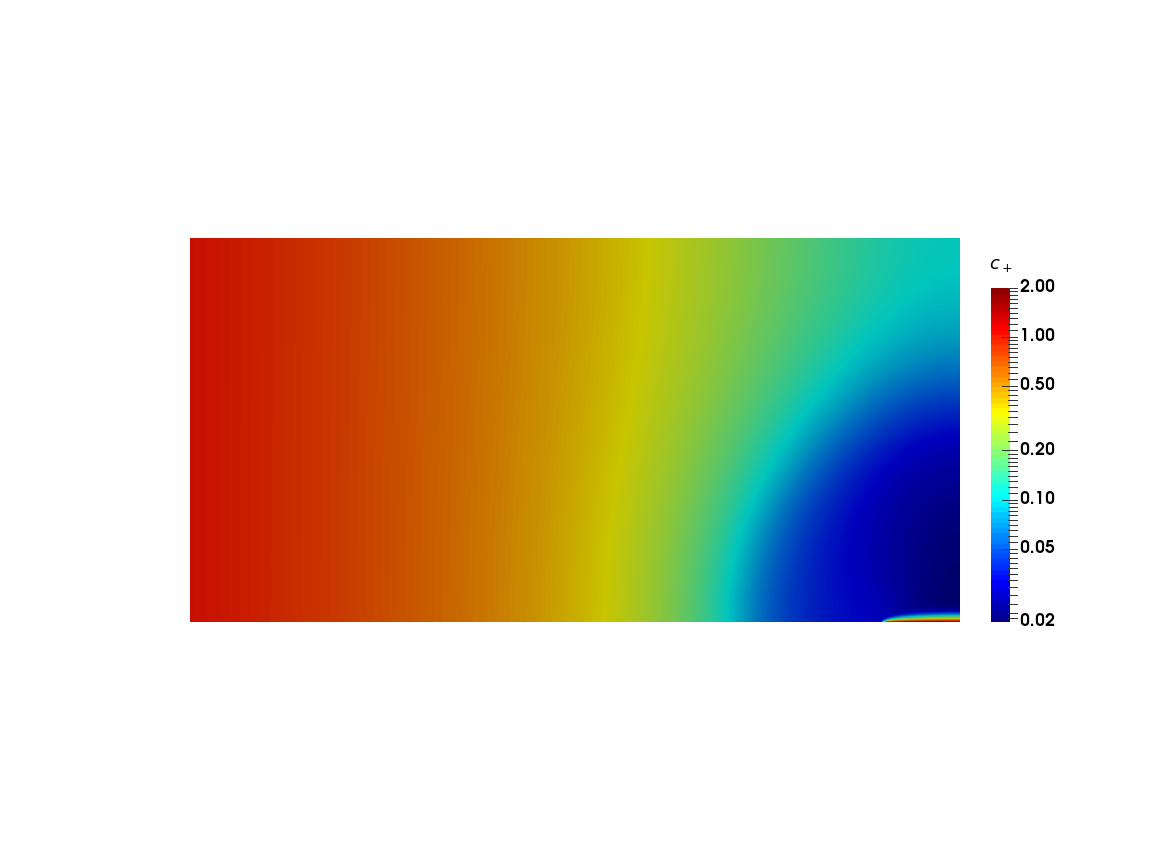}
\caption{\(c_+\) concentration near the cation selective membrane.}
\label{fig:2DWeakComp}
\end{figure}

To simulate the formation of an IDZ, the Poisson-Nernst-Planck equations were solved for the left half of the anodic channel. The boundary condition for the cation concentration is strongly enforced at the inlet (\(c_+=1\)) and weakly at the membrane (\(c_+=2\)). The concentration of the anion is set to \(c_-=1\) at the inlet. The boundary condition defining potential is strongly enforced at the inlet (\(\phi=50\)) and weakly at the membrane (\(\phi=0\)). At the walls and the line of symmetry, a no flux boundary condition (\(\boldsymbol{j}_i\cdot\boldsymbol{n} = 0\)) was applied. We use an unstructured triangular mesh that exhibits moderate mesh refinement at the membrane and a coarse mesh close to the inlet. A contour plot of cation concentration is shown in Figure \ref{fig:2DWeakComp}. Notice that, at the membrane boundary, a thin cation boundary layer is formed; along with the formation of the IDZ. 

The boundary flux of cation \(c_+\) at the membrane was calculated from Eq. (\ref{eq:flxCal}) and compared with the calculation from strong BC, for progressively refined meshes (that are refined close to the membrane). Like the results from 1D, the boundary flux calculation obtained by using weak BC show remarkable convergence even for coarse mesh resolutions. This effect is clearly seen in Figure~\ref{fig:2Dflx}, which shows that the weakly imposed boundary is able to accurately capture the flux even for fairly coarse meshes. 

\usetikzlibrary{pgfplots.groupplots}

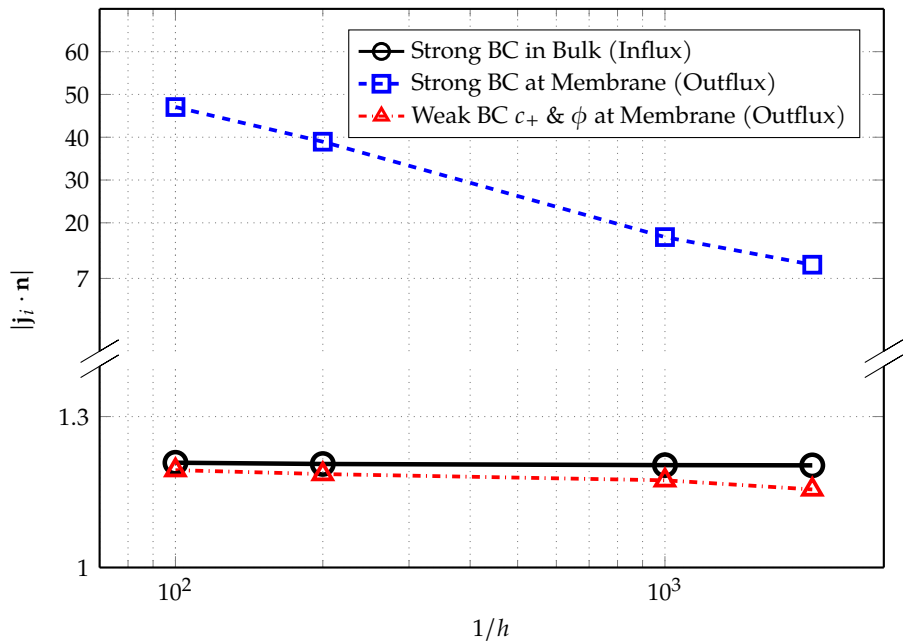
\begin{figure}
    \centering
    \pgfplotsset{
    every non boxed x axis/.style={} 
}
    \begin{tikzpicture}
    \tikzstyle{every node}=[font=\footnotesize]
\begin{groupplot}[
    group style={
        group name=my fancy plots,
        group size=1 by 2,
        xticklabels at=edge bottom,
        vertical sep=2mm
    },
    width=0.7 \linewidth,
    xmin=100,
    xmax=2000,
    xmode=log,
    grid=both,
    grid style={black!50,dotted},
    axis line style = thick,
    legend style={at={(0.97,0.94)},anchor=north east}, 
]

\nextgroupplot[axis x line=top,
            legend cell align={left},
               ymin=-10,
               ymax=70,
               ytick = {7, 20, 30, 40, 50, 60},
               height=0.36 \linewidth,
              xmin = 70,
              xmax = 2800,
               y label style={at={(axis description cs:0.05,0.15)},anchor=south},
               ylabel=\(|\vec{j}_{i}\cdot\vec{n}|\)]
        \addplot[black, 
                mark = o, 
                mark size=3pt, 
                mark options={solid}, line width=0.4mm]
            table[x expr={\thisrow{inv_h}},y expr={-100*\thisrow{strIn}},col sep=space]{5.RESULTS/data/2Dflx.txt};
        \addplot[blue, dashed, 
                mark = square, 
                mark size=3pt, 
                mark options={solid}, line width=0.5mm]
           table[x expr={\thisrow{inv_h}},y expr={\thisrow{strOut}},col sep=space]{5.RESULTS/data/2Dflx.txt};
        \addplot[red, dash dot, 
                mark = triangle, 
                mark size=3pt, mark options={solid}, line width=0.4mm]
            table[x expr={\thisrow{inv_h}},y expr={-100*\thisrow{WeakCP}},col sep=space]{5.RESULTS/data/2Dflx.txt};
     
        \legend{{Strong BC in Bulk (Influx)}, {Strong BC at Membrane (Outflux)}, {Weak BC $c_+$ \& $\phi$ at Membrane (Outflux)}}
        \coordinate(dl)at(rel axis cs:0,0);
        \coordinate(dl2)at(rel axis cs:1,0);
        
\nextgroupplot[axis x line=bottom,
              height=0.25 \linewidth,
              xlabel=$1/h$,
              ytick = {1, 1.3},
               xmin = 70,
               xmax = 2800,
              ymin = 1.0,ymax = 1.4]
        \addplot[black, mark = o, mark size=4pt, mark options={solid}, line width=0.6mm]
            table[x expr={\thisrow{inv_h}},y expr={\thisrow{strIn}},col sep=space]{5.RESULTS/data/2Dflx.txt};
        \addplot[red, dash dot, mark = triangle, mark size=4pt, mark options={solid}, line width=0.5mm]
            table[x expr={\thisrow{inv_h}},y expr={\thisrow{WeakCP}},col sep=space]{5.RESULTS/data/2Dflx.txt};
            
        \coordinate(ul)at(rel axis cs:0,1);
        \coordinate(ul2)at(rel axis cs:1,1);
            
\end{groupplot}
\draw(dl)--+(0.25,0.15)--+(-0.25,-0.15);
\draw(dl2)--+(0.25,0.15)--+(-0.25,-0.15);
\draw(ul)--+(0.25,0.15)--+(-0.25,-0.15);
\draw(ul2)--+(0.25,0.15)--+(-0.25,-0.15);

\end{tikzpicture}
    \caption{Boundary flux calculation at the membrane with weak and strong boundary conditions, and their comparison with influx at bulk.}
    \label{fig:2Dflx}
\end{figure}

\subsection{Electrokinetic analyte separation: 3D simulation} 
\begin{figure}[!hbtp]
\begin{subfigure}{.5\textwidth}
    \centering
    \includegraphics[width = \linewidth]{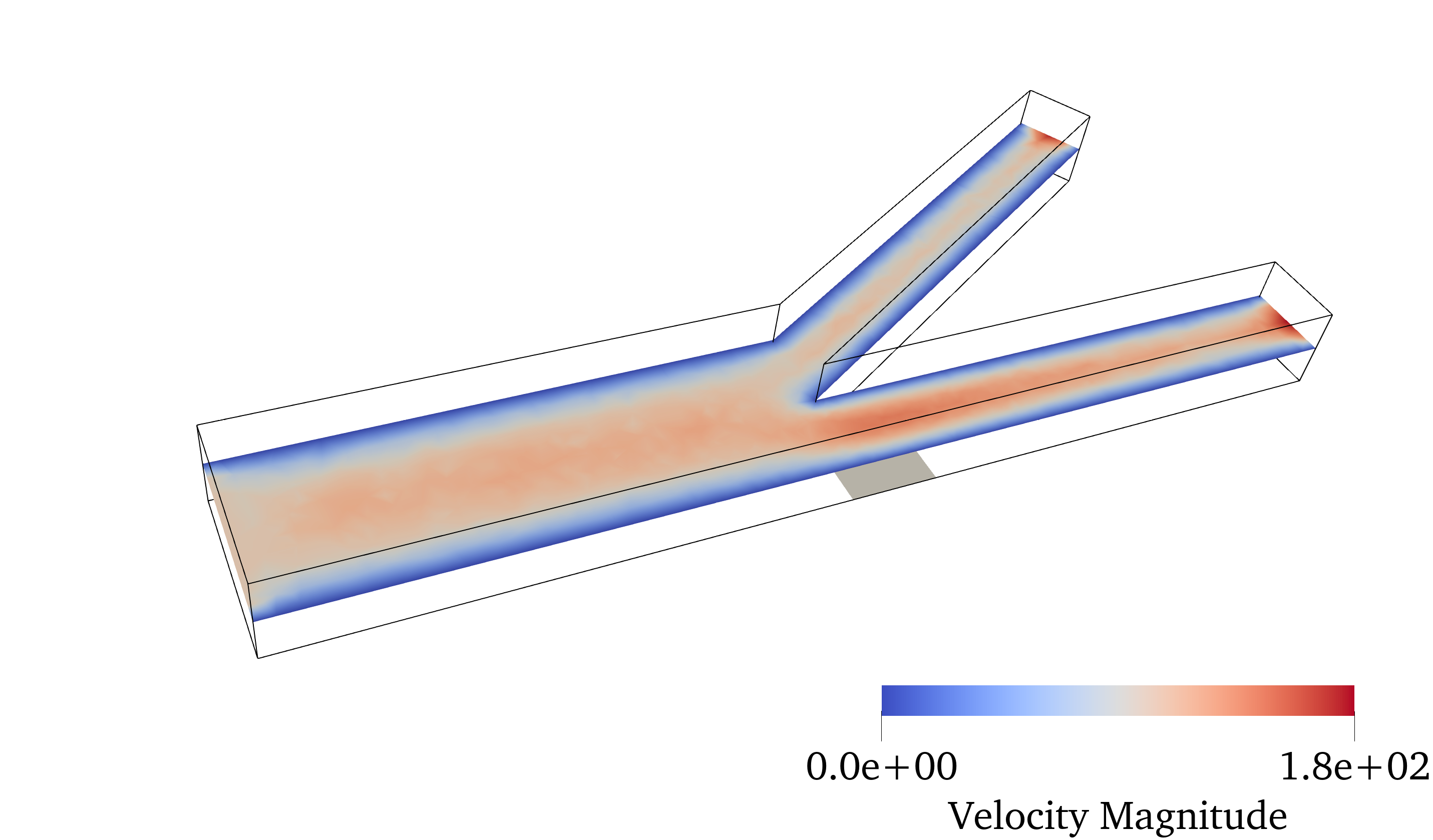}
    \subcaption{}
\end{subfigure}
\begin{subfigure}{.5\textwidth}
    \centering
    \includegraphics[width = \linewidth]{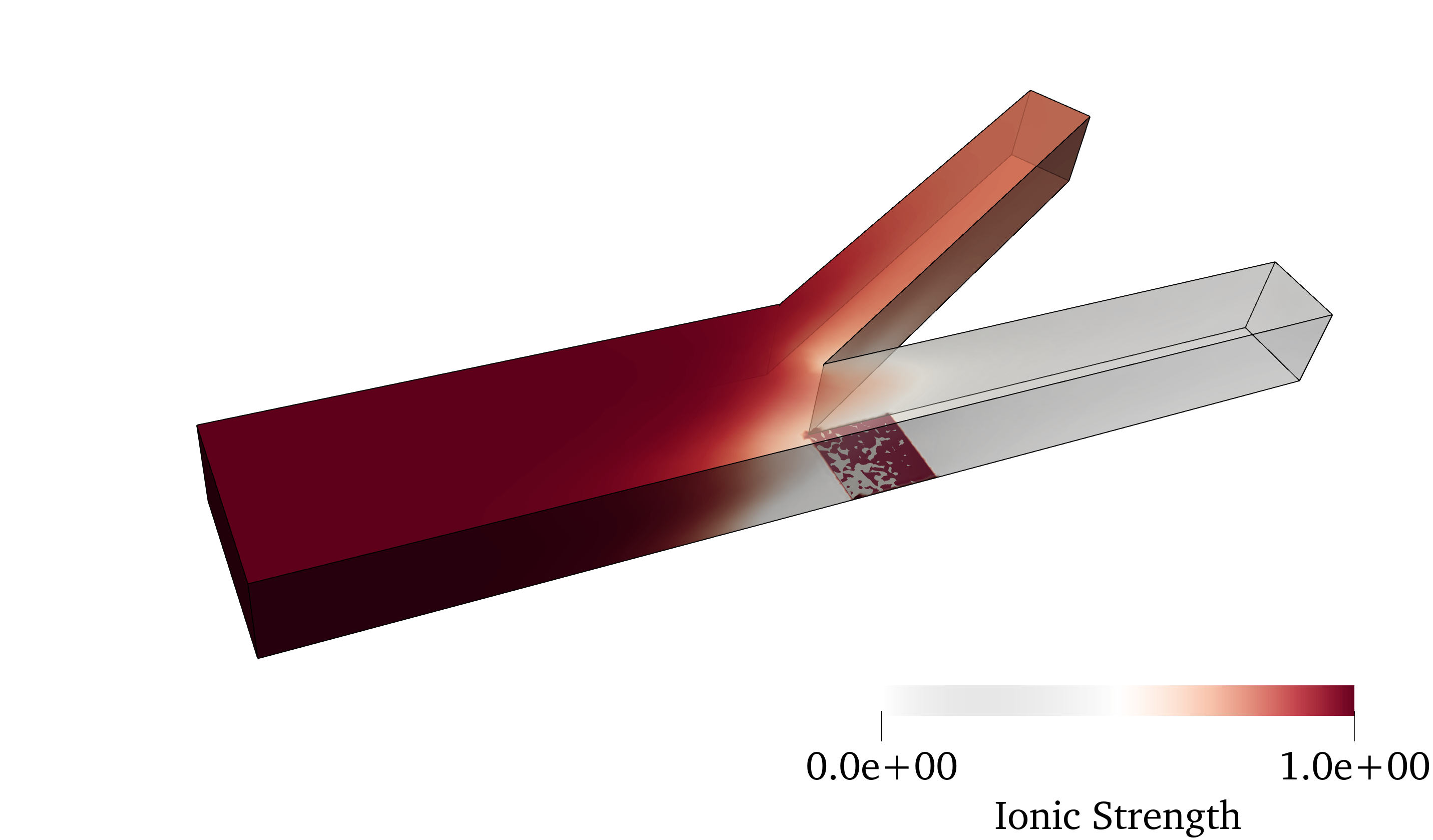}
    \subcaption{}
\end{subfigure}
\newline
\begin{subfigure}{.5\textwidth}
    \centering
    \includegraphics[width = \linewidth]{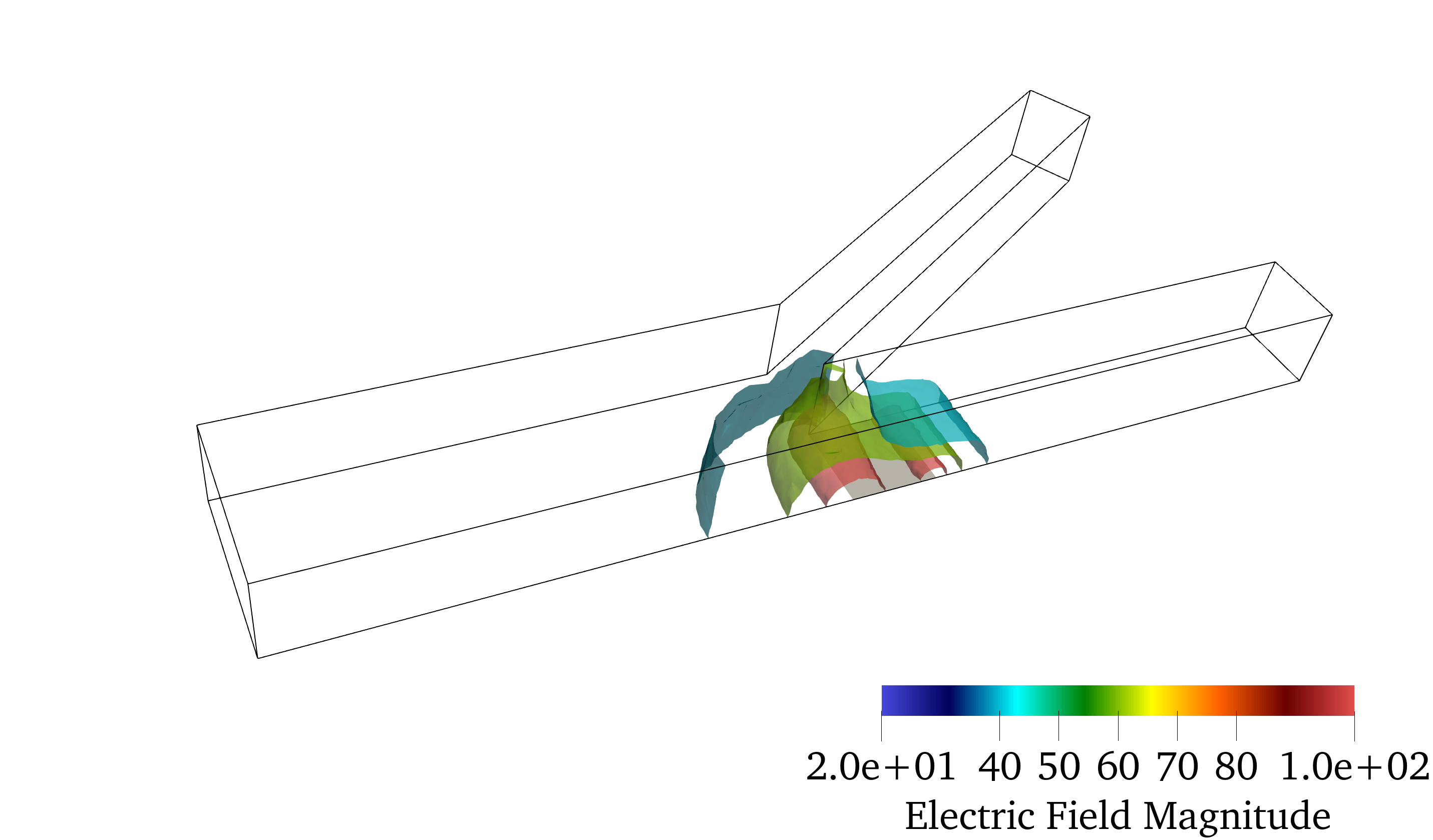}
    \subcaption{}
\end{subfigure}
\begin{subfigure}{.5\textwidth}
    \centering
    \includegraphics[width = \linewidth]{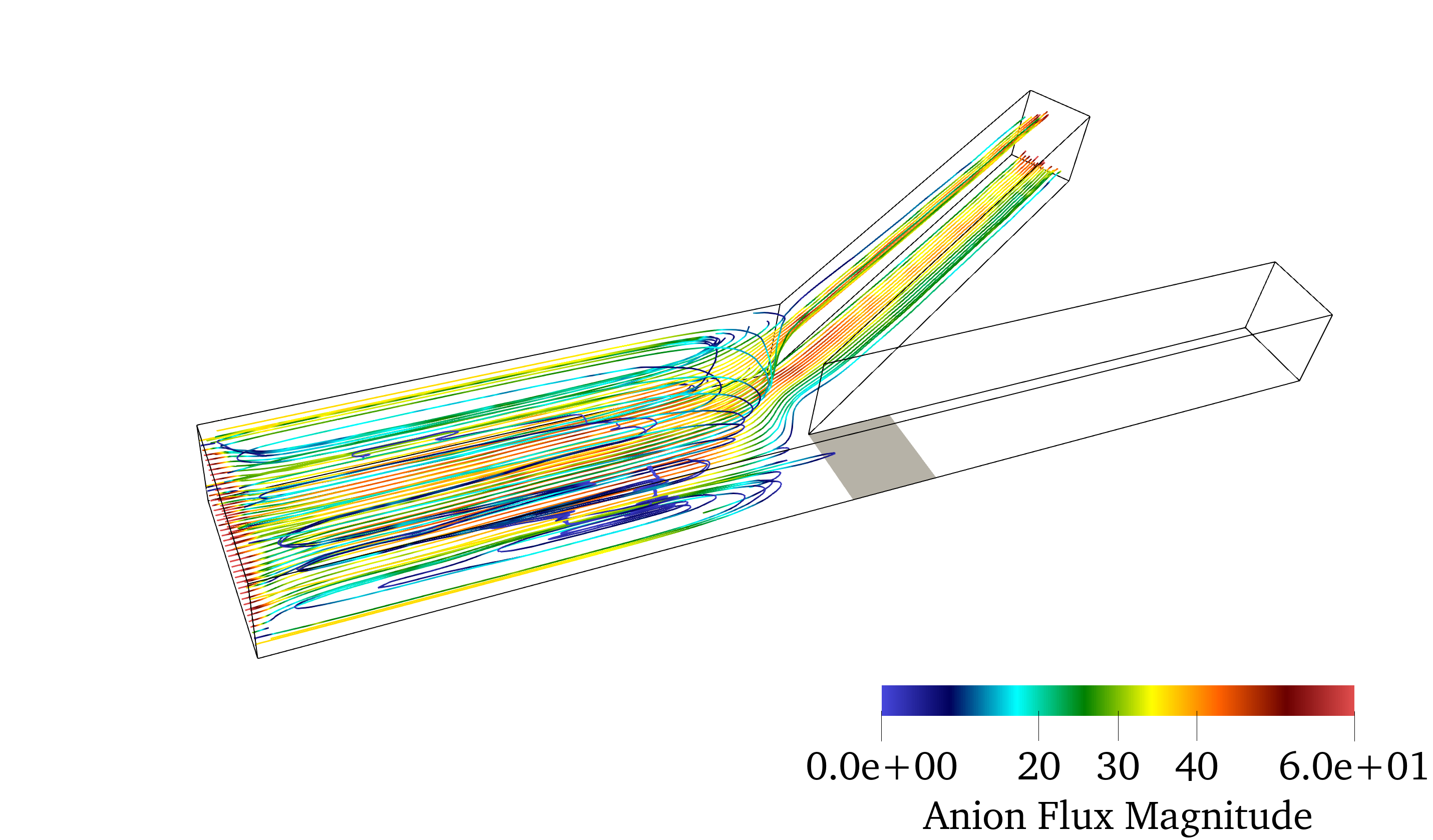}
    \subcaption{}
\end{subfigure}
\caption{Analyte separation in a diverging microfluidic channel. (a) Magnitude of flow velocity. (b) ionic strength, $I=1/2\sum_{i=1}^{N} c_i z_i^{2}$, which is an average concentration of species for binary electrolyte. (c) Iso-surfaces of electric field near the membrane. The membrane is shown in gray at the bottom surface after the junction. (d) Anion flux is screened out from the membrane channel by high electric field.}
\label{fig:3DSeparation}
\end{figure}

In this section, we test our platform on a canonical electrochemical system --- electrolyte separation (desalting) in a 3D microchannel equipped with a permselective membrane. 
The device configuration is shown in Figure \ref{fig:3DSeparation}, with the channel branching into two channels. The cation selective membrane is located along the outer wall of the straight channel, just downstream of the branch point. 
A constant flow of electrolyte is maintained by a pressure difference, Figure \ref{fig:3DSeparation} (a). A potential difference across the microchannel is applied, with anodic conditions applied at the inlet, and ground conditions applied at the membrane surface. The fully coupled Navier-Stokes Poisson-Nernst-Planck equations were solved for the microchannel geometry. An unstructured tetrahedral mesh was created using the mesh generating software, Gmsh (V2.10.1). The non-dimensional parameters defining this system are as follows: \(Sc=686.81\), \(\kappa=0.39\), and \(\Lambda=0.097\). 

The boundary conditions for the variables at various boundaries are as follows:
(a) Membrane surface: weak imposition of Dirichlet conditions for potential ($\phi = 0$), and cation concentration ($c_+ = 2$), and zero anion current flux boundary condition (\(\vec{j_{-}}\cdot \vec{n}=0\)); (b) Inlet: Strong imposition of Dirichlet condition for potential ($\phi = 150$), anion concentration ($c_+=1$), cation concentration ($c_{-}=1$), and inlet velocity ($u=100$); (c) Walls: no flux boundary conditions for both species, no slip for velocity; (d) Outlet: pressure set to zero. 

The steady state results are shown in Figure \ref{fig:3DSeparation}. 
Once the electric field is applied, an IDZ forms near the membrane surface, as can be seen in Figure \ref{fig:3DSeparation} (b). This dramatic reduction in the concentration of the conductive species upstream of the perm-selective membrane creates a high electric field at the junction of the splitting channels as can be seen in Figure \ref{fig:3DSeparation} (c). This electric field, in conjunction with the pressure driven flow results in separation of the electrolyte. Specifically, as charged species are transported to the channel by bulk flow, anions entering the lower channel are screened by the high electric field and are redirected to the upper channel, Figure \ref{fig:3DSeparation} (d). Conversely, cations are attracted by the electric field and removed out of the device through the cation selective membrane. This results in near complete removal of charged species from the channel that is intersected by the junction. This result illustrates the mechanism used in water purification and fluid management related to hemodialysis~\cite{kim2010direct, knust2013electrochemically}.

To showcase the impact of weak imposition of boundary conditions, we check how well total flux is conserved. That is, we compute the sum of cation fluxes across the inlet, outlet and membrane, which following conservation of charge should sum to zero. Since we expect sharp gradients of concentrations near the membrane, we vary the discretization near the membrane while keeping the mesh size in the rest of the domain fixed at \(\SI{3e-2}{}\). Table \ref{Tab:flux_3D} compares this total flux in the case of strong imposition of boundary conditions at the membrane versus weak imposition of boundary conditions at the membrane. We can clearly see that the weak imposition of boundary conditions results in significantly more accurate fluxes, even for relatively coarse meshes.


As before, we note that the flux calculation at the membrane remains challenging due to the large gradients of variables. This result is important, as it shows that even complex 3D electrochemical systems can be efficiently simulated using weakly imposed boundary conditions.  

\begin{table}
\centering
\begin{tabular}{c|cc} 
\toprule
\(h_{mem}\)   &   Strong BC   &   Weak BC     \\
\hline\hline
\SI{6e-2}{}         &   7.92    &  -1.73   \\
\SI{3e-2}{}         &   11.9      &   -0.9 \\
\SI{1.5e-2}{}       &   12.8     &   -0.631 \\
\SI{7.5e-3}{}      &   12.8     &  -0.638 \\
\bottomrule
\end{tabular}
\caption{Net cation flux at inlet, outlet, and membrane for controlled mesh size at membrane, \(h_{mem}\). Inlet fluid velocity was \(u = 100\)}.
\label{Tab:flux_3D}
\end{table}

\begin{figure}[!hbtp]
\begin{subfigure}{.5\textwidth}
    \centering
    \includegraphics[width = \linewidth]{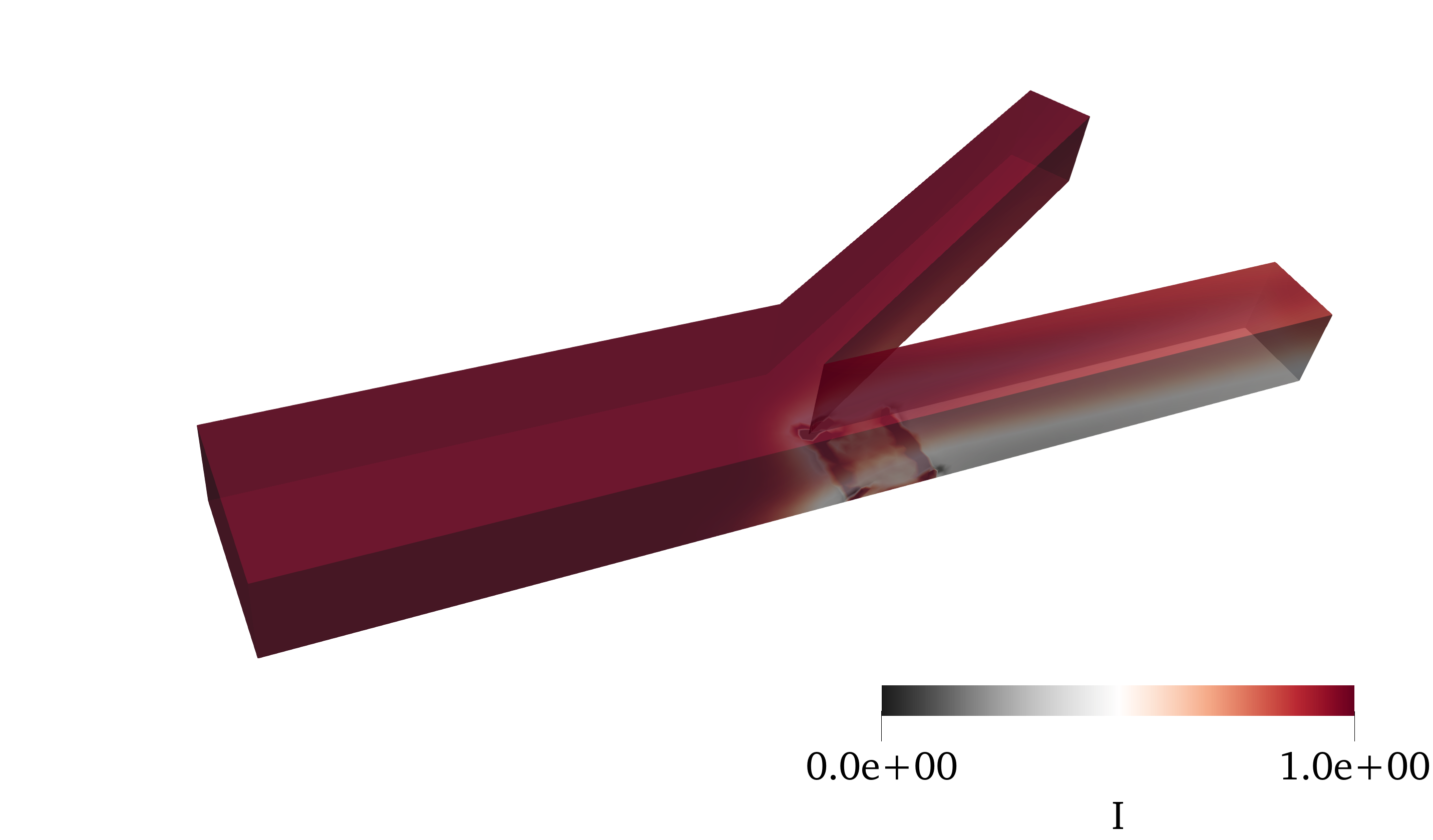}
    \subcaption{}
\end{subfigure}
\begin{subfigure}{.5\textwidth}
    \centering
    \includegraphics[width = \linewidth]{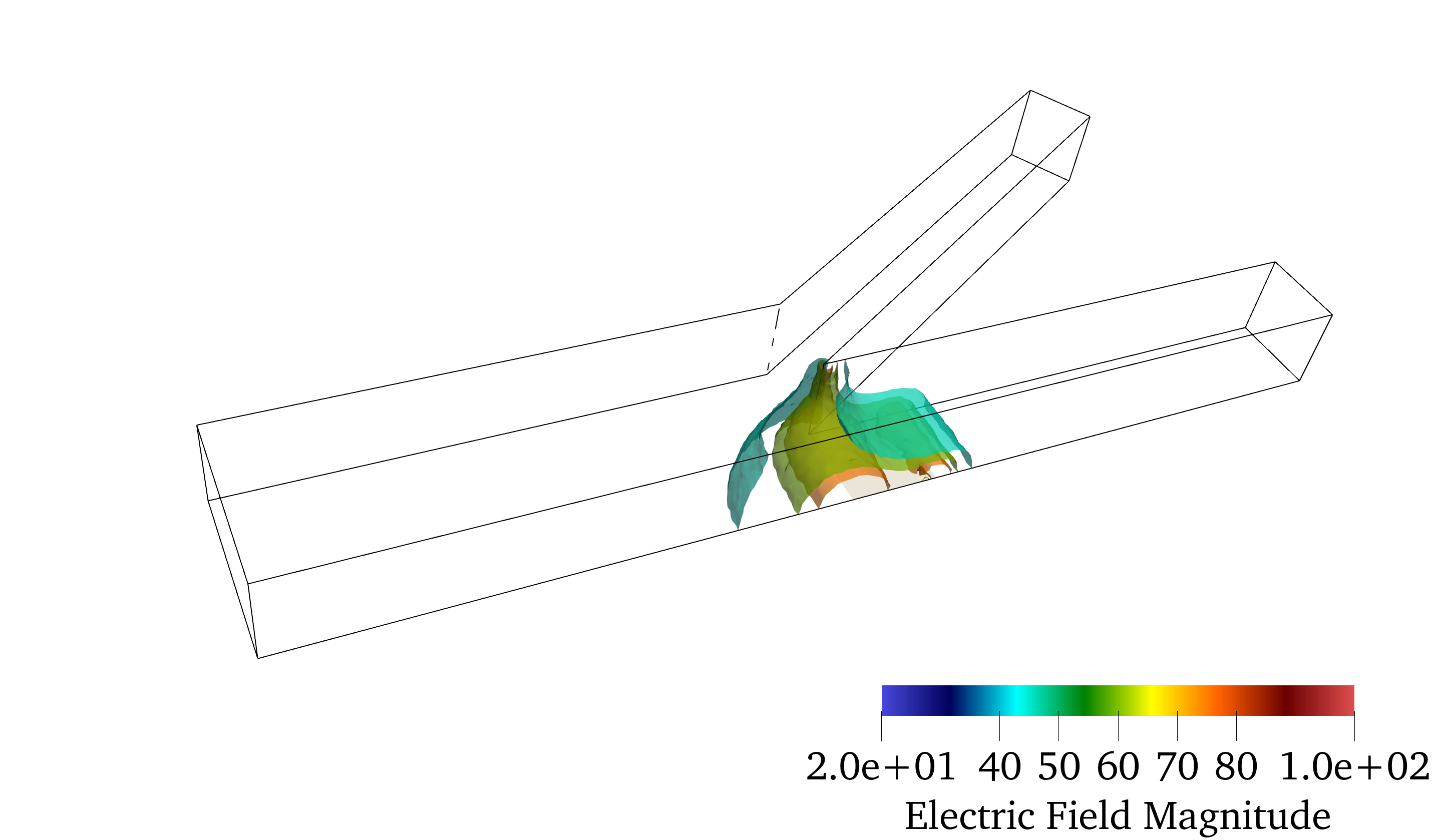}
    \subcaption{}
\end{subfigure}
\newline
\begin{subfigure}{.5\textwidth}
    \centering
    \includegraphics[width = \linewidth]{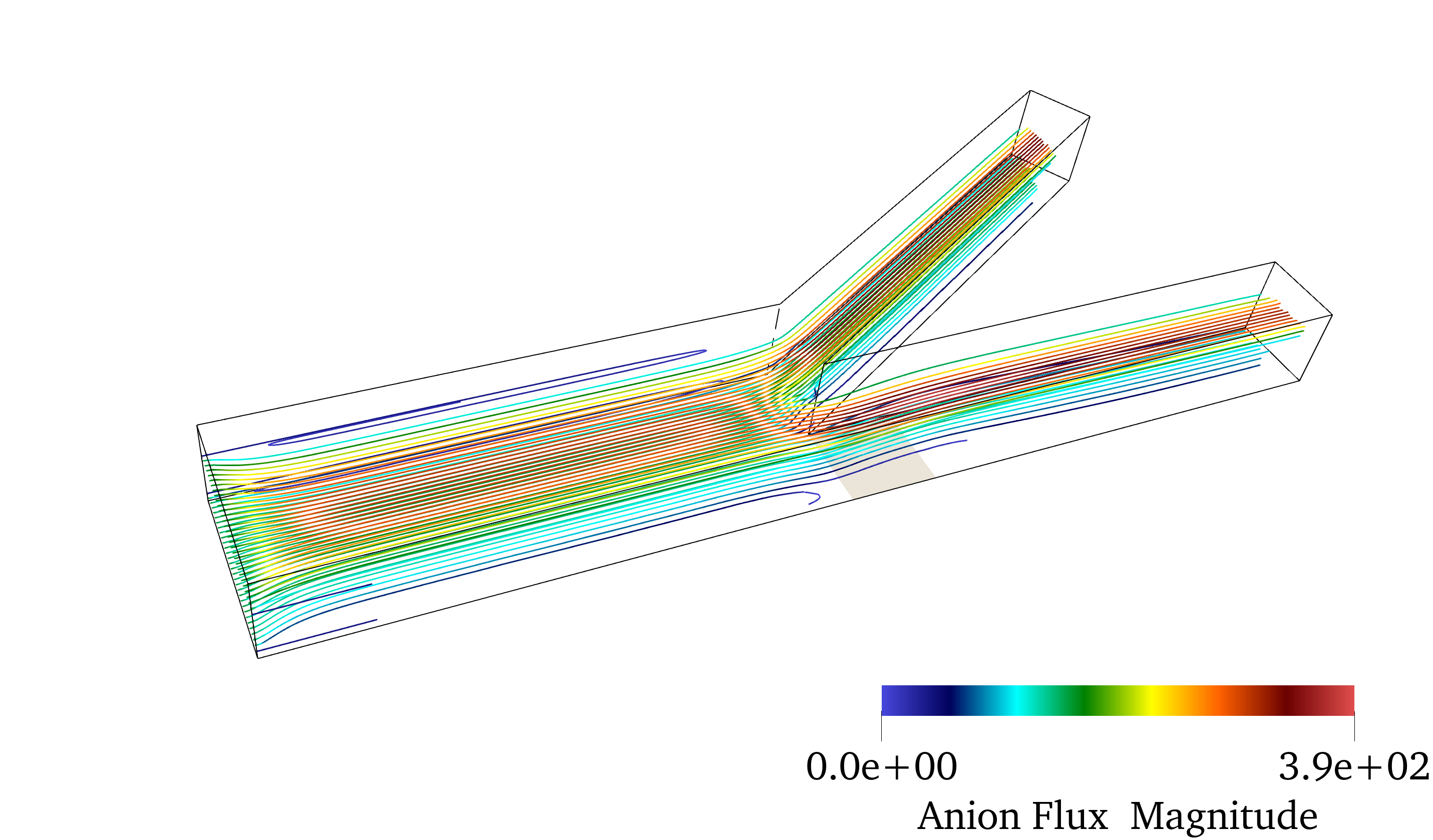}
    \subcaption{}
\end{subfigure}
\begin{subfigure}{.5\textwidth}
    \centering
    \includegraphics[width = \linewidth]{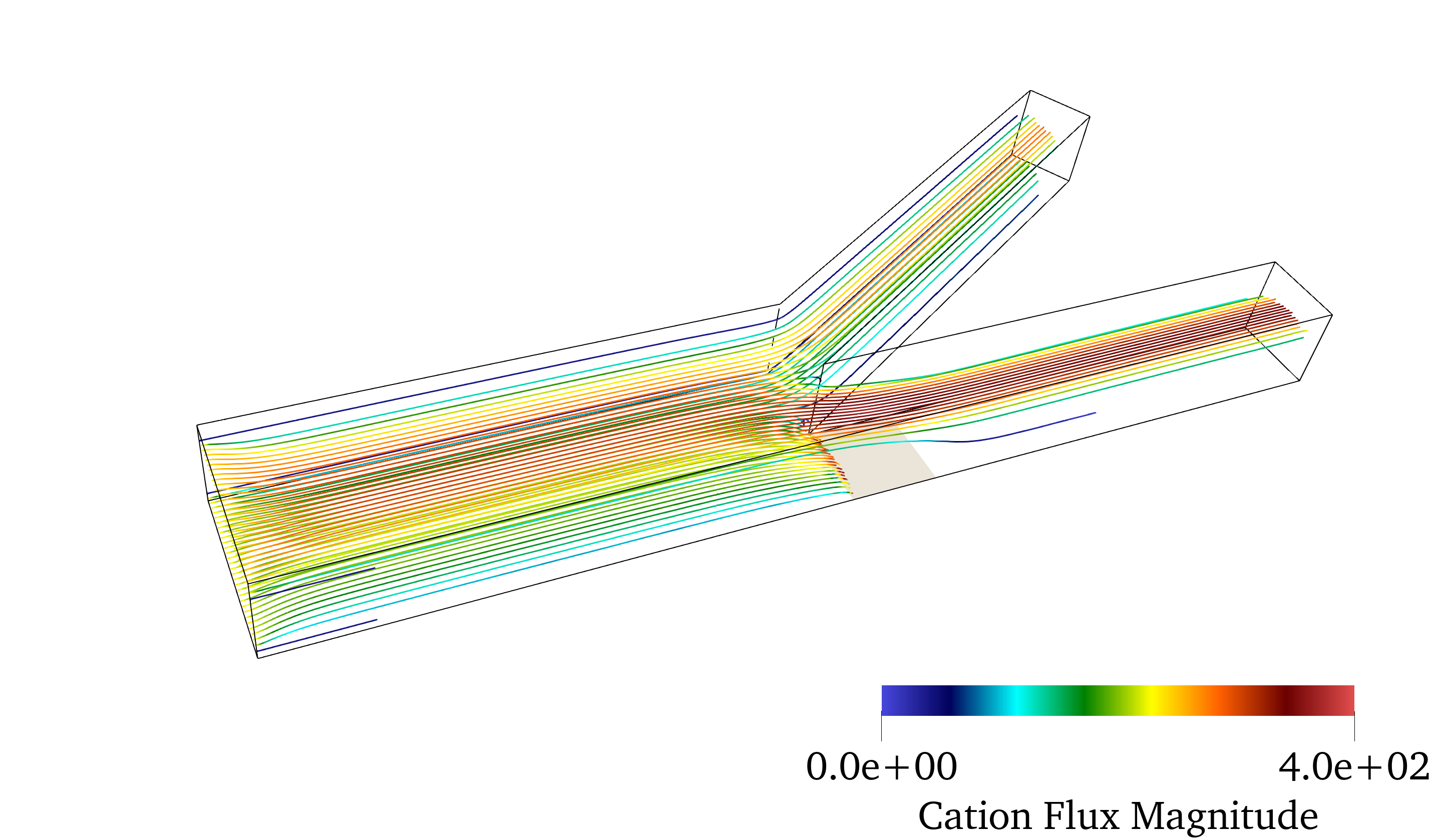}
    \subcaption{}
\end{subfigure}
\caption{ \textit{Failure to achieve separation during high influx:} Analyte separation in the diverging channel with excessive influx. When the convection is dominant over the electric field, the separation at the junction is not maintained. (a) ionic strength, $I=1/2\sum_{i=1}^{N} c_i z_i^{2}$, which is an average concentration of species for binary electrolyte; (b) Iso-surfaces of electric field near the membrane. The membrane is shown in gray at the bottom surface after the junction. (c) Anion flux passes through the membrane channel due to high convection; (d) Cation flux either passes through the membrane region or leaves through the membrane.}
\label{fig:3DSeparationU200}
\end{figure}

The previous numerical example showed electrochemical separation as a consequence of a balance between the electric migration away from the IDZ and the convective flux due to imposed flow rate. We conclude this section by simulating a case where the imposed flow rate is sufficiently large to overwhelm the electromigration, thus disrupting the separation process. Figure \ref{fig:3DSeparationU200} shows the result obtained when the flow rate is doubled relative to the previous case, while the potential and inlet concentration remains unchanged. Figure \ref{fig:3DSeparationU200}(a) show high ionic strength at both outlets, which implies high cation and anion concentration at both outlets indicating a failure to achieve separation. The structure and magnitude of electric field around the membrane is similar to that observed in the earlier simulation, see iso-surfaces in Figure \ref{fig:3DSeparationU200} (b). However, due to high convection, the charged species penetrate the electric field barrier at the channel junction, see Figure \ref{fig:3DSeparationU200} (c) and (d). As a result, a significant amount of cation and anion concentration is found at the membrane outlet as well.  

We again evaluate the conservation of total cation flux in the simulation. Table \ref{Tab:flux_3D_U200} shows the net flux across the permselective membrane calculated for a range of mesh sizes. The net flux resulting from the weak BC approach is significantly closer to zero, indicating a clear advantage in comparison to a strong imposition of boundary conditions.

\begin{table}
\centering
\begin{tabular}{c|cc} 
\toprule
\(h_{mem}\)   &   Strong BC   &   Weak BC     \\
\hline\hline
\SI{6e-2}{}         &   3.50    &  -1.43   \\
\SI{3e-2}{}         &   7.74      &   -0.75  \\
\SI{1.5e-2}{}       &   8.84     &   -0.46 \\
\SI{7.5e-3}{}      &   8.86     &  -0.46 \\
\bottomrule
\end{tabular}
\caption{Net cation flux at inlet, outlet, and membrane for controlled mesh size at membrane, \(h_{mem}\). Inlet fluid velocity was \(u = 100\)}.
\label{Tab:flux_3D_U200}
\end{table}

\subsection{Electrokinetic instability near a perm selective membrane} 
In this sub-section we illustrate the ability of the framework to capture electroconvective instability. This instability occurs due to the interplay between the hydrodynamics with electrostatic forces~\citep{druzgalski2013direct} causing chaotic fingers of charge density to emanate from an ion selective membrane (beyond a critical applied electric field). This is an interesting, yet challenging phenomena to capture, requiring very fine resolution close to the membrane boundary. Here, we show that qualitatively identical results can be simulated using relatively coarse meshes having no more than 2 elements across the Debye layer. 

We consider a long rectangular channel with an aspect ratio $8\times1$. The cation selective membrane boundary conditions are enforced at the bottom, with reservoir boundary conditions enforced at the top. A potential difference of $120$ is maintained across the domain. This corresponds to (a) cation boundary conditions of $C_{+} = 2$ at the bottom and $C_{+}=1$ at the top, (b) anion boundary conditions of zero flux at the bottom, and $C_{-}=1$ at the top, (c) potential boundary conditions of 120 at the top and 0 at the bottom, (d) no slip boundary conditions for velocity at the top and bottom. Symmetric boundary conditions are applied to the side walls. 

This domain is discretized using a rectangular mesh with 1280 $\times$ 180 quad elements created using the Gmsh (V2.10.1) software. A geometric progression ($\texttt{ratio} = 1.021$) based stretching was applied along the height to get a clustered mesh. This produced a mesh refined at the bottom, with the smallest element exhibiting a height of $\sim \SI{5e-4}{}$. 

The non-dimensional Debye layer, $\Lambda$ is $\SI{1e-3}{}$ thick. 
This corresponds to a little less than two elements representing the Debye layer in the mesh. 
Note that this scenario is a realistic case corresponding to $\sim \SI{10}{\micro\metre}$ channel heights seen in several electrolyte applications~\citep{kim2010direct, berzina2020tutorial,kim2010nanofluidic}. 
The Schmidt number, $Sc$, is $\SI{1e-3}{}$ and the electrohydrodynamic coupling constant,$\kappa$, is 0.5. The Dirichlet-Neumann transition for the cation is applied at the bottom by weakly enforcing the concentration. 
A time step of $\Delta t = \SI{1e-6}{}$ is used to solve this problem. 

\begin{figure}[!hbtp]
\centering
\includegraphics[clip,width=\textwidth]{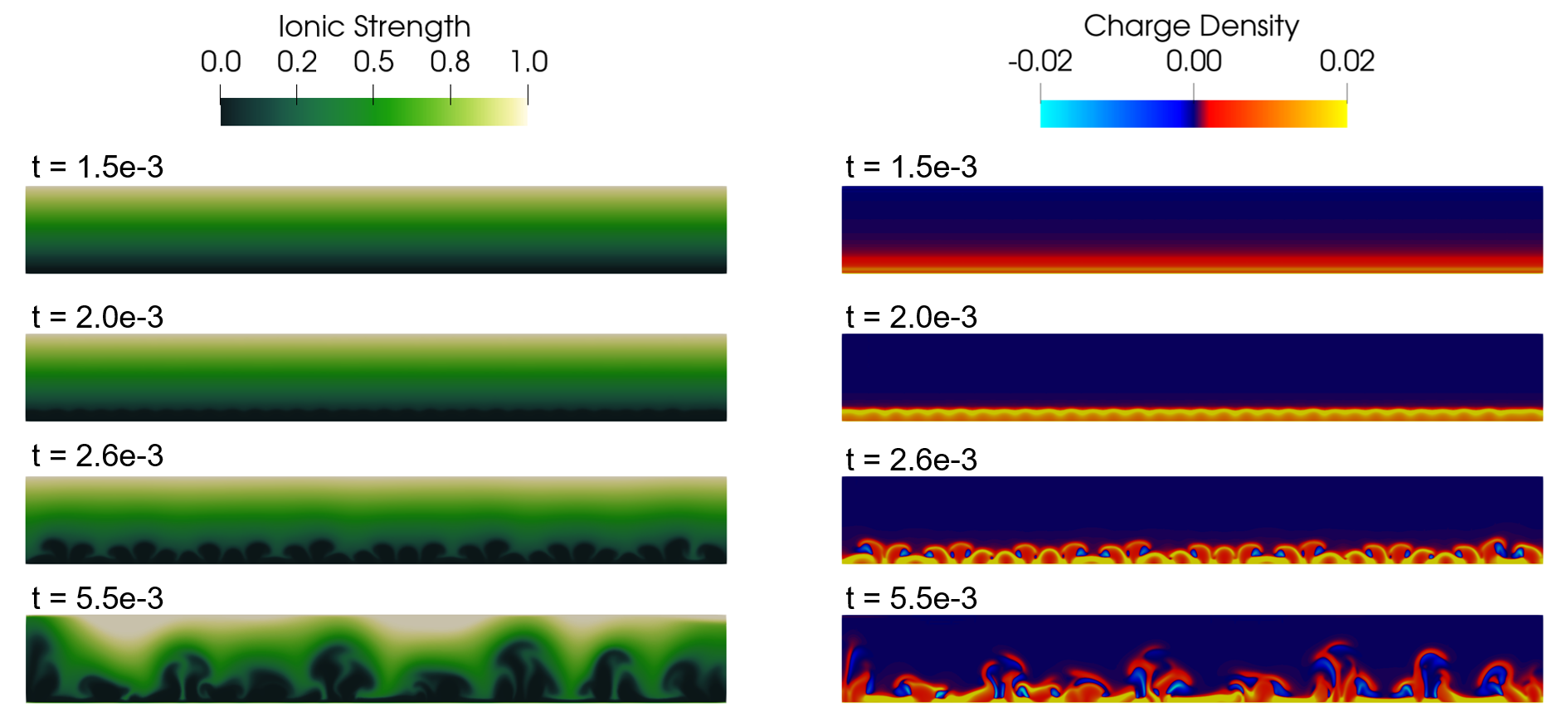}
    \caption{Electrokinetic instabilities near the cation selective membrane. Ionic strength (left) and charge density (right). $\Lambda = \SI{1e-3}{}$ at various timepoints ($t$ is non-dimensional time)}
\label{fig:instability}
\end{figure}

Figure \ref{fig:instability} shows the development of the instability in the system, matching the results from benchmark simulations~\cite{druzgalski2013direct}. The results at initial times are analogous to the 1D results seen in section \ref{1D_IDZ}. The electrolyte concentration is stratified with the formation of a depletion zone near the membrane. Note the charge separation in the depletion zone. This non-zero charge density together with the local electric field drives the fluid flow, which results in the electrokinetic instabilities at later times. These appear as fingers of differential charge density emanating from the ion selective membrane. This qualitatively matches the numerical and analytical studies  reported in \cite{druzgalski2013direct, zaltzman2007electro} and experimental studies in \cite{kwak2013microscale, yossifon2008selection}. We defer a more qualitative comparison of these simulations (including a parametric sweep across various potential differences) to a subsequent study. 


\section{Conclusion}\label{sec:conclusion}
In this study, we simulate electrokinetic systems represented by the Navier-Stokes-Poisson-Nernst-Planck equations, with a key quantity of interest being the current flux at the system boundaries. Accurately computing the current flux is challenging due to the thin boundary layers (small Debye lengths) that require fine mesh to resolve.  We address this challenge by using the Dirichlet-to-Neumann transformation to weakly impose the Dirichlet conditions. The framework was validated against manufactured solutions and the analytical solution for electroosmotic flow. We next simulated the dynamics near a permselective membrane in 1D and 2D. Then, the simulation was tested in a 3D application for electrolyte separation (desalting) in a branching microchannel. Lastly, the electrokinetic instability near a perm-selective membrane was simulated by the coupled Navier-Stokes-Poisson-Nernst-Planck equations. 
We showed that weak imposition of boundary conditions can produce accurate boundary flux values, even with a coarse mesh and independent of flow conditions. This approach substantially reduces the computational cost of modeling complex electrochemical systems.  

\section{Acknowledgments}
We thank Kumar Saurabh from the Ganapathysubramanian  group for technical discussions and implementation support, as well as for proof-reading the manuscript. 


\section*{References}
\bibliographystyle{model1-num-names.bst}
\bibliography{bibfile.bib}

\appendix
\section{Different types of time scaling}
\label{app:lambdaT}
 The smallest timescales in electrochemical system are electric double layer charging time and chemical reaction time. In the current work frame, chemical reaction is not considered; thus, the double layer charging time is the smallest time scale of the system. The double layer charging time is directly correlated with the double layer thickness in steady state \cite{morrow2006time}. Therefore, it is reasonable to consider the double layer thickness as a term in defining the characteristic timescale,
 \begin{equation}
     \tau=\frac{L\lambda}{D}.
 \end{equation}{}
The corresponding non-dimensional Nernst-Planck equation becomes 
\begin{equation}\label{eq:NDNP}
    \frac{dc_{i}}{dt} + \Lambda \boldsymbol{u} \cdot \nabla c_{i}
    = \Lambda \nabla \cdot (\nabla c_{i} + z_{i} c_{i}\nabla \phi), 
\end{equation}
the Poisson equation
\begin{equation}\label{eq:NDP}
 -2 \Lambda^2 \nabla^2\phi = \rho_e,
\end{equation}
and the Navier-Stokes equation
\begin{equation}\label{eq:NDNS}
    \frac{1}{Sc}\frac{1}{\Lambda}\frac{d\boldsymbol{u}}{dt} + \frac{1}{Sc} \boldsymbol{u}\cdot\nabla\boldsymbol{u}
        =-\nabla p + \nabla^{2}\boldsymbol{u}
        - \frac{\kappa}{2\Lambda^2}\Sigma c_i z_i\nabla \phi
\end{equation}

\section{Additional results}
\label{app:WeakDEP}
We provide additional results illustrating cation distribution comparisons between strong and weak imposition of boundary conditions in ~\ref{fig:2DWeakComp}. We focus on the near boundary region ($0 \le x \le 0.1$), for the problem setup discussed in \ref{1D_IDZ} with \(\Lambda=0.01\). The simulation for the strong boundary conditions is performed on a mesh with $1000$ uniform elements ($h = 10^{-3}$) in the domain. Thus, there are about 10 elements across the boundary layer. In contrast, we use fairly coarse meshes with $100$ and $80$ uniform elements for simulations with weak imposition of the boundary conditions. A single element in this mesh is comparable to the boundary layer thickness.
As described in the main text, two types of weak BC are considered: Weak BC for both \(c+\) and \(\phi\), and weak BC only for \(c+\). We note that in both these cases (and meshes) the current flux matches very well (as shown in the main text). It is interesting to see that even with coarse meshes, the cation distribution matches with the highly resolved cation distribution within two elements from the boundary. 


\begin{figure}[b!]
\centering
\begin{subfigure}[t]{0.45\linewidth}
\centering
\begin{tikzpicture}[]
        \tikzstyle{every node}=[font=\footnotesize]
      \begin{axis}[
          width=\linewidth, 
          y label style={at={(axis description cs:0.1,0.5)},anchor=south},
          xlabel=$x$, 
          ylabel=$c_+$,
          legend style={at={(0.95,0.95)},anchor=north east}, 
          x tick label style={rotate=0,anchor=north}, 
          axis line style = thick,
          cycle list name=color list,
          legend cell align={left},
          xmin=0,
          xmax=0.1
        ]
        \addplot[black, solid, line width=0.6mm]
        table[x expr={\thisrow{x}},y expr={\thisrow{cp}},col sep=space]{5.RESULTS/data/1DdepWeak/Str.txt};
        
        \addplot[blue, dashed, line width=0.5mm, mark = o, mark size=4pt, mark options={solid}]
        table[x expr={\thisrow{x}},y expr={\thisrow{cp}},col sep=space]{5.RESULTS/data/1DdepWeak/PC100.txt};
        
        \addplot[red, dash dot, line width=0.5mm, mark = x, mark size=4pt, mark options={solid}]
        table[x expr={\thisrow{x}},y expr={\thisrow{cp}},col sep=space]{5.RESULTS/data/1DdepWeak/C100.txt};
        
        \legend{Strong BC,Weak BC \(c_+\) \& \(\phi\),Weak BC \(c_+\)}
        \end{axis}
\end{tikzpicture}
\caption{100 elements}
\end{subfigure}
\begin{subfigure}[t]{0.45\linewidth}
\centering
\begin{tikzpicture}[]
\tikzstyle{every node}=[font=\footnotesize]
\begin{axis}[
          width=\linewidth, 
          y label style={at={(axis description cs:0.1,0.5)},anchor=south},
          xlabel=$x$, 
          ylabel=$c_+$,
          legend style={at={(0.95,0.95)},anchor=north east}, 
          x tick label style={rotate=0,anchor=north}, 
          axis line style = thick,
          cycle list name=color list,
          legend cell align={left},
          xmin=0,
          xmax=0.1
        ]
        \addplot[black, solid, line width=0.6mm]
        table[x expr={\thisrow{x}},y expr={\thisrow{cp}},col sep=space]{5.RESULTS/data/1DdepWeak/Str.txt};
        
        \addplot[blue, dashed, line width=0.5mm, mark = o, mark size=4pt, mark options={solid}]
        table[x expr={\thisrow{x}},y expr={\thisrow{cp}},col sep=space]{5.RESULTS/data/1DdepWeak/PC80.txt};
        
        \addplot[red, dash dot, line width=0.5mm, mark = x, mark size=4pt, mark options={solid}]
        table[x expr={\thisrow{x}},y expr={\thisrow{cp}},col sep=space]{5.RESULTS/data/1DdepWeak/C80.txt};
        
        \legend{Strong BC,Weak BC \(c_+\) \& \(\phi\),Weak BC \(c_+\)}
        \end{axis}
\end{tikzpicture}
\caption{80 elements}
\end{subfigure}

\caption{1D depletion: comparison between weakly imposed BC both for \(c_+\) and \(\phi\) and only for \(c_+\). Note that only adjacent of the membran ($x<0.1$) is shown.}
\label{fig:1DWeakComp}
\end{figure}
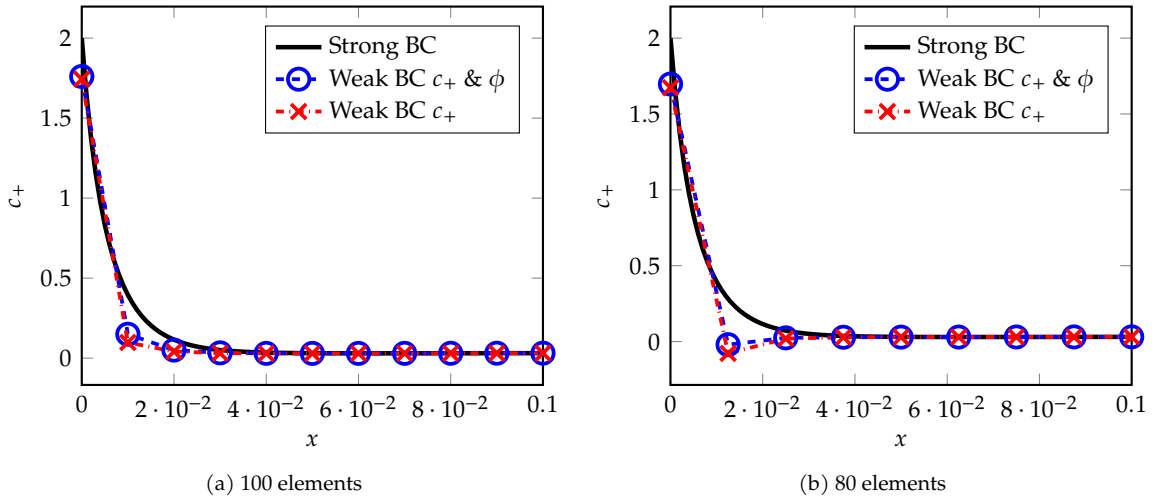







\end{document}